# Conventional and unconventional anomalous velocities in multiband systems

Katsuhiko Higuchi[1,*], Wakano Sakamoto[2] and Masahiko Higuchi[2,**]

[1] Graduate School of Advanced Science and Engineering, Hiroshima University, Higashi-Hiroshima 739-8527, Japan

[2] Department of Physics, Faculty of Science, Shinshu University, Matsumoto 390-8621, Japan

*khiguchi@hiroshima-u.ac.jp

**higuchi@shinshu-u.ac.jp

## ABSTRACT

The anomalous velocity has been derived so far based on the single-band approximation. In this paper, the anomalous velocity is derived accounting for multiple energy bands. It is shown that when multiple energy bands are considered, the anomalous velocity is actually derived from the velocity term which goes to zero under the single-band approximation. It is also shown that the anomalous velocity based on the single-band approximation is derived improperly from the velocity term which becomes zero in considering multiple energy bands. Furthermore, it is found that unconventional types of anomalous velocity may appear in addition to the conventional anomalous velocity. These unconventional anomalous velocities are perpendicular to the electric field and come from the singularity of the magnetic Bloch function in the magnetic Brillouin zone. It is confirmed that conventional and unconventional anomalous velocities can also be derived not only from the steady-state perturbation theory but also from the time-dependent perturbation theory in which a time-dependent vector potential yielding the uniform electric field is treated as a perturbation.

## I. INTRODUCTION

The electronic and magnetic properties of materials can be discussed by investigating the dynamics of a wave function composed of the Bloch functions [1-15]. For example, if an external electric field is applied to a material, and if the wave function of electron is supposed to be described by the linear combination of the Bloch functions with a single band index, then the velocity of the electron is given as the sum of the group velocity and the anomalous velocity [5, 6]. It is well-known that the anomalous velocity is responsible for the quantum anomalous Hall effect [7-13]. Also, the de Haas-van Alphen (dHvA) effect can be described by using a semiclassical equation of motion for the Bloch electron [14-16]. Furthermore, the semiclassical equation of motion has been used in the analysis and design of electronic devices [17-22].

The electronic state of a material immersed in a uniform magnetic field is described by the magnetic Bloch function [23]. The magnetic Bloch state is classified by a wavevector lying in the magnetic first Brillouin zone (MBZ) [23,24]. Such the energy band structure described in the MBZ is sometimes referred to as the magnetic Bloch band. In recent years, we have developed a method to calculate the magnetic Bloch bands [24-29]. By analysing the magnetic Bloch bands, we have successfully revisited the dHvA effect [25] and the oscillation phenomenon associated with magnetic breakdown [26]. Furthermore, we have predicted additional oscillations of magnetization [25] that could not be explained by the conventional Lifsitz–Kosevich theory [14].

The quantized Hall conductivity is also described by means of the magnetic Bloch function [30-40]. By using the Kubo formula with the magnetic Bloch function, it is shown that the Hall conductivity is given in terms of the anomalous velocity that is expressed as the Berry curvature of the magnetic Bloch state [30]. Then, the integral of the anomalous velocity over wavevectors lying in the MBZ is shown to be equal to an integer called the Chern number [13,30,31]. Therefore, the Hall conductivity, which is proportional to the sum of the Chern number with respect to magnetic Bloch bands, is shown to be quantized if the Fermi energy lies between magnetic Bloch bands [13,30,31]. Note that the Fermi energy lies between magnetic Bloch bands due to a disorder potential [41-43]. One of the key points for describing the quantized Hall conductivity by means of magnetic Bloch states is that the MBZ can be divided into multiple patches, and the phase of the magnetic Bloch function is smooth within each patch, but discontinuous at the boundaries between patches as shown by Kohmoto [31]. Because of this phase discontinuity of the magnetic Bloch function, the Chern number, which is expressed as an integral of the Berry curvature of the magnetic Bloch band, is known to have a nonzero integer value [31]. In other words, the phase discontinuity of the magnetic Bloch function plays an essential role when describing the quantized Hall conductivity by means of magnetic Bloch states [31].

On the other hand, in previous works [4,9-13], the anomalous velocity is derived under the assumption that the phase of the Bloch function (or the magnetic Bloch function) is locally continuous around the wavepacket momentum varying in the BZ (or MBZ). Due to this assumption, the surface integral over the surface of the BZ (or MBZ), which appears in the derivation process of the anomalous velocity, is treated as zero. Recently, the existence of additional anomalous velocities has been studied in the field of mathematical physics for solid-state phenomena [44-47]. To the extent that the single-band approximation is valid, it is shown that one gets the equation of motion that contains not only the conventional anomalous velocity but also additional velocity terms [44-47]. Thus, although anomalous velocity has been extensively studied, the effects of multiple energy bands and phase discontinuities of magnetic Bloch functions on anomalous velocity have not been thoroughly discussed.

In this paper, we investigate the dynamics of the electron moving in materials immersed in a magnetic field with an applied electric field, accounting for both the multiple energy bands and the phase discontinuity of the magnetic Bloch function. The resultant anomalous velocity terms, which are perpendicular to the electric field, consists of two types. One type corresponds to the conventional anomalous velocity. The other is regarded as an unconventional anomalous velocity and is caused by the phase discontinuity of the magnetic Bloch function. In addition, comparing the derivation process of anomalous velocities in considering multiple energy bands with that under the single-band approximation, it will be shown that the correct origin of the anomalous velocity is different from that in the single-band approximation. We also show that the conventional and unconventional anomalous velocities can be derived, irrespective of whether the uniform electric field is represented as a scalar potential or as a time-dependent vector potential.

## II. EXPRESSION OF VELOCITY FOR MAGNETIC BLOCH FUNCTION

The Hamiltonian for an electron under both the periodic potential $V(\boldsymbol{r})$ of a crystal and the vector potential $\boldsymbol{A}(\boldsymbol{r})$ of a uniform magnetic field is given by

$$\hat{H}_0 = \frac{1}{2m}\left\{\boldsymbol{p} + e\boldsymbol{A}(\boldsymbol{r})\right\}^2 + V(\boldsymbol{r}) . \tag{1}$$

If the Landau gauge is employed for $\boldsymbol{A}(\boldsymbol{r})$, then $\boldsymbol{A}(\boldsymbol{r})$ is given by $(0, Bx, 0)$. Here we neglect the electron spin. The eigenfunction of $\hat{H}_0$ is referred to as the magnetic Bloch function that is denoted by $\phi_{n\boldsymbol{\kappa}}(\boldsymbol{r})$, where the subscripts $n$ and $\boldsymbol{\kappa}$ denote the index of the magnetic Bloch band and the wavevector lying within the MBZ, respectively [23,24]. The Schrödinger equation is given by

$$\hat{H}_0\phi_{n\boldsymbol{\kappa}}(\boldsymbol{r}) = \varepsilon_n(\boldsymbol{\kappa})\phi_{n\boldsymbol{\kappa}}(\boldsymbol{r}), \tag{2}$$

where $\varepsilon_n(\boldsymbol{\kappa})$ is the eigenvalue. Let us consider the case where a uniform electric field $\boldsymbol{E}$ is applied to the system. The Hamiltonian of the system is given by

$$\hat{H} = \hat{H}_0 - \boldsymbol{F}\cdot\boldsymbol{r}, \tag{3}$$

where $\boldsymbol{F} = -e\boldsymbol{E}$. We shall comment on the scalar potential $-\boldsymbol{F}\cdot\boldsymbol{r}$ that yields a uniform electric field. Since this scalar potential $-\boldsymbol{F}\cdot\boldsymbol{r}$ is incompatible with the periodic boundary conditions of the magnetic Bloch function and includes the position operator $\boldsymbol{r}$, it may cause difficulties such as the possibility of $\boldsymbol{r}$ going to infinity [48] and the choice of the coordinate origin. As mentioned in Sec. III, Appendices A and B, such difficulties can be avoided by following the previous works [2-4, 7, 12, 49-53]. Note that an alternative way to treat a uniform electric field is to introduce a uniform and time-dependent vector potential [11, 48, 54, 55]. We will discuss the alternative way to treat a uniform electric field in Sec. IV.

The time-dependent wavefunction $\psi(\boldsymbol{r},t)$ is supposed to be constructed by superposing magnetic Bloch functions, i.e.,

$$\psi(\boldsymbol{r},t) = \sum_{n\boldsymbol{\kappa}} a_n(\boldsymbol{\kappa},t)\phi_{n\boldsymbol{\kappa}}(\boldsymbol{r}), \tag{4}$$

where $a_n(\boldsymbol{\kappa},t)$ denotes the expansion coefficient and satisfies $\sum_{n\boldsymbol{\kappa}}|a_n(\boldsymbol{\kappa},t)| = 1$ due to the normalization condition of $\psi(\boldsymbol{r},t)$. Here note that the summation is done over wavevectors in the MBZ and band indexes. The expectation value of $\boldsymbol{r}$ with respect to $\psi(\boldsymbol{r},t)$ is

$$\langle\boldsymbol{r}\rangle = \sum_{n\boldsymbol{\kappa}}\sum_{n'\boldsymbol{\kappa}'} a_n^*(\boldsymbol{\kappa},t)a_{n'}(\boldsymbol{\kappa}',t)\int_\Omega \phi_{n\boldsymbol{\kappa}}^*(\boldsymbol{r})\boldsymbol{r}\phi_{n'\boldsymbol{\kappa}'}(\boldsymbol{r})d^3r. \tag{5}$$

To investigate the velocity of the magnetic Bloch electron, we consider the derivative of $\langle \boldsymbol{r} \rangle$ with respect to $t$. Substitution of Eq. (4) into the time-dependent Schrödinger equation leads to the set of equations for $a_n(\boldsymbol{\kappa},t)$ and $a_n^*(\boldsymbol{\kappa},t)$:

$$\begin{aligned} i\hbar \frac{\partial a_n(\boldsymbol{\kappa},t)}{\partial t} &= \varepsilon_n(\boldsymbol{\kappa}) a_n(\boldsymbol{\kappa},t) - \sum_{n'\boldsymbol{\kappa}'} a_{n'}(\boldsymbol{\kappa}',t) \int_\Omega \phi^*_{n\boldsymbol{\kappa}}(\boldsymbol{r})(\boldsymbol{F}\cdot\boldsymbol{r})\phi_{n'\boldsymbol{\kappa}'}(\boldsymbol{r}) d^3r, \\ -i\hbar \frac{\partial a^*_n(\boldsymbol{\kappa},t)}{\partial t} &= \varepsilon_n(\boldsymbol{\kappa}) a^*_n(\boldsymbol{\kappa},t) - \sum_{n'\boldsymbol{\kappa}'} a^*_{n'}(\boldsymbol{\kappa}',t) \int_\Omega \phi_{n\boldsymbol{\kappa}}(\boldsymbol{r})(\boldsymbol{F}\cdot\boldsymbol{r})\phi^*_{n'\boldsymbol{\kappa}'}(\boldsymbol{r}) d^3r. \end{aligned} \tag{6}$$

Note that the matrix elements of $\boldsymbol{r}$ with respect to the magnetic Bloch functions appear in the right-hand side of Eqs. (5) and (6). Details of calculating the matrix elements will be discussed in the next section, Appendices A and B. By using Eqs. (5) and (6), we get the expression for the derivative of $\langle \boldsymbol{r} \rangle$ with respect to $t$:

$$\begin{aligned} \frac{d}{dt}\langle \boldsymbol{r} \rangle = {} & \frac{1}{i\hbar}\sum_{n\boldsymbol{\kappa}}\sum_{n'\boldsymbol{\kappa}'} a^*_n(\boldsymbol{\kappa},t) a_{n'}(\boldsymbol{\kappa}',t) \left\{ \varepsilon_{n'}(\boldsymbol{\kappa}') - \varepsilon_n(\boldsymbol{\kappa}) \right\} \int_\Omega \phi^*_{n\boldsymbol{\kappa}}(\boldsymbol{r}) \boldsymbol{r} \phi_{n'\boldsymbol{\kappa}'}(\boldsymbol{r}) d^3r \\ & - \frac{1}{i\hbar}\sum_{n\boldsymbol{\kappa}}\sum_{n'\boldsymbol{\kappa}'} a^*_n(\boldsymbol{\kappa},t) a_{n'}(\boldsymbol{\kappa}',t) \sum_{n''\boldsymbol{\kappa}''} \Big\{ \int_\Omega \phi^*_{n\boldsymbol{\kappa}}(\boldsymbol{r}') \boldsymbol{r}' \phi_{n''\boldsymbol{\kappa}''}(\boldsymbol{r}') d^3r' \int_\Omega \phi^*_{n''\boldsymbol{\kappa}''}(\boldsymbol{r}) \left(\boldsymbol{F}\cdot\boldsymbol{r}\right) \phi_{n'\boldsymbol{\kappa}'}(\boldsymbol{r}) d^3r \\ & \qquad - \int_\Omega \phi^*_{n\boldsymbol{\kappa}}(\boldsymbol{r}') \left(\boldsymbol{F}\cdot\boldsymbol{r}'\right) \phi_{n''\boldsymbol{\kappa}''}(\boldsymbol{r}') d^3r' \int_\Omega \phi^*_{n''\boldsymbol{\kappa}''}(\boldsymbol{r}) \boldsymbol{r} \phi_{n'\boldsymbol{\kappa}'}(\boldsymbol{r}) d^3r \Big\}. \end{aligned} \tag{7}$$

Adopting the completeness of the magnetic Bloch functions,

$$\sum_{n''\boldsymbol{\kappa}''} \phi^*_{n''\boldsymbol{\kappa}''}(\boldsymbol{r}) \phi_{n''\boldsymbol{\kappa}''}(\boldsymbol{r}') = \delta(\boldsymbol{r}-\boldsymbol{r}'), \tag{8}$$

to the second term in the right-hand side of Eq. (7), then the second term vanishes. Therefore, we obtain

$$\frac{d}{dt}\langle \boldsymbol{r} \rangle = \frac{1}{i\hbar}\sum_{n\boldsymbol{\kappa}}\sum_{n'\boldsymbol{\kappa}'} \left\{ \varepsilon_{n'}(\boldsymbol{\kappa}') - \varepsilon_n(\boldsymbol{\kappa}) \right\} a^*_n(\boldsymbol{\kappa},t) a_{n'}(\boldsymbol{\kappa}',t) \int_\Omega \phi^*_{n\boldsymbol{\kappa}}(\boldsymbol{r}) \boldsymbol{r} \phi_{n'\boldsymbol{\kappa}'}(\boldsymbol{r}) d^3r. \tag{9}$$

It should be noted that $d\langle \boldsymbol{r} \rangle / dt$ does not directly depend on electric field but depends on the electric field only through the expansion coefficient. This is the most significant difference from the single-band approximation. That is, if we adopt the single-band approximation, the second term of the right-hand side of Eq. (7) which directly depends on the electric filed would remain. This can be confirmed easily as follows. Under the single-band approximation, the wave function is supposed to be constructed by magnetic Bloch functions with a single magnetic Bloch band $n$.

$$\psi_n(\boldsymbol{r},t)=\sum_{\boldsymbol{\kappa}} a(\boldsymbol{\kappa},t)\phi_{n\boldsymbol{\kappa}}(\boldsymbol{r}), \tag{10}$$

where $a(\boldsymbol{\kappa},t)$ denotes the expansion coefficient and satisfies $\sum_{\boldsymbol{\kappa}}|a(\boldsymbol{\kappa},t)|=1$. Note that Eq. (10) is an approximation. Similarly to Eq. (9), we have the following expression of the velocity under the single-band approximation:

$$\begin{aligned}\frac{d}{dt}\langle \boldsymbol{r}\rangle_n^{\mathrm{SB}}=&\frac{1}{i\hbar}\sum_{\boldsymbol{\kappa}}\sum_{\boldsymbol{\kappa}'}a^*(\boldsymbol{\kappa},t)a(\boldsymbol{\kappa}',t)\left\{\varepsilon_n(\boldsymbol{\kappa}')-\varepsilon_n(\boldsymbol{\kappa})\right\}\int_\Omega \phi_{n\boldsymbol{\kappa}}^*(\boldsymbol{r})\boldsymbol{r}\phi_{n\boldsymbol{\kappa}'}(\boldsymbol{r})d^3r\\ &-\frac{1}{i\hbar}\sum_{\boldsymbol{\kappa}}\sum_{\boldsymbol{\kappa}'}a^*(\boldsymbol{\kappa},t)a(\boldsymbol{\kappa}',t)\sum_{\boldsymbol{\kappa}''}\left\{\int_\Omega \phi_{n\boldsymbol{\kappa}}^*(\boldsymbol{r}')\boldsymbol{r}'\phi_{n\boldsymbol{\kappa}''}(\boldsymbol{r}')d^3r'\int_\Omega \phi_{n\boldsymbol{\kappa}''}^*(\boldsymbol{r})\left(\boldsymbol{F}\cdot\boldsymbol{r}\right)\phi_{n\boldsymbol{\kappa}'}(\boldsymbol{r})d^3r\right.\\ &\left.-\int_\Omega \phi_{n\boldsymbol{\kappa}}^*(\boldsymbol{r}')\left(\boldsymbol{F}\cdot\boldsymbol{r}'\right)\phi_{n\boldsymbol{\kappa}''}(\boldsymbol{r}')d^3r'\int_\Omega \phi_{n\boldsymbol{\kappa}''}^*(\boldsymbol{r})\boldsymbol{r}\phi_{n\boldsymbol{\kappa}'}(\boldsymbol{r})d^3r\right\},\end{aligned} \tag{11}$$

where $\langle \boldsymbol{r}\rangle_n^{\mathrm{SB}}$ denotes the expectation value with respect to the wave function $\psi_n(\boldsymbol{r},t)$. Comparing Eq. (11) with Eq. (9), the first term of Eq. (11) corresponds to Eq. (9). The second term of Eq. (11) does not vanish under the single-band approximation, while the corresponding second term of Eq. (7) correctly vanishes due to Eq. (8) in considering multiple energy bands as mentioned above. Therefore, the remaining second term of Eq. (11) is regarded to come from the insufficient description of the wave function given by Eq. (10). As shown in the next section, the second term of Eq. (11) accidentally leads to the conventional anomalous velocity with several additional terms, while the conventional and unconventional anomalous velocities will be derived from Eq. (9) correctly.

## III. CONVENTIONAL AND NOVEL ANOMALOUS VELOCITIES

Let us rewrite the right-hand side of Eqs. (9) and (11). For this aim, the matrix elements of $\boldsymbol{r}$ with respect to the magnetic Bloch functions should be evaluated avoiding the difficulties associated with the position operator $\boldsymbol{r}$ [46]. Following previous works [2-4, 7, 12, 47-51], we employ the magnetic Bloch theorem to avoid this problem. First, the magnetic Bloch function is formally expressed by

$$\phi_{n\boldsymbol{\kappa}}(\boldsymbol{r}) = e^{i\boldsymbol{\kappa}\cdot\boldsymbol{r}} u_{n\boldsymbol{\kappa}}(\boldsymbol{r}) . \tag{12}$$

According to the magnetic Bloch theorem [23,24], the function $u_{n\boldsymbol{\kappa}}(\boldsymbol{r})$ in Eq. (12) should satisfy the following relation;

$$u_{n\boldsymbol{\kappa}}(\boldsymbol{r} - \boldsymbol{t}_\alpha) = e^{-i\frac{e}{\hbar}\chi(\boldsymbol{r},\boldsymbol{t}_\alpha)} u_{n\boldsymbol{\kappa}}(\boldsymbol{r}) , \tag{13}$$

where $\boldsymbol{t}_\alpha$ denotes a translation vector such that magnetic translation operators for $\boldsymbol{t}_\alpha{}'s$ commute with each other [23,24]. The function $\chi(\boldsymbol{r},\boldsymbol{t}_\alpha)$ in the right-hand side of Eq. (13) is defined by $\boldsymbol{A}(\boldsymbol{r} - \boldsymbol{t}_\alpha) = \boldsymbol{A}(\boldsymbol{r}) + \nabla\chi(\boldsymbol{r},\boldsymbol{t}_\alpha)$. If the Landau gauge is employed for $\boldsymbol{A}(\boldsymbol{r})$, then $\chi(\boldsymbol{r},\boldsymbol{t}_\alpha)$ is given by $-Bt_{\alpha x} y$ [24]. By using the relation $\nabla_{\boldsymbol{\kappa}'} e^{i\boldsymbol{\kappa}'\cdot\boldsymbol{r}} = i\boldsymbol{r} e^{i\boldsymbol{\kappa}'\cdot\boldsymbol{r}}$ and Eqs. (12) and (13), the matrix element of $\boldsymbol{r}$ with respect to the magnetic Bloch states can be rewritten as

$$\int_\Omega \phi^*_{n\boldsymbol{\kappa}}(\boldsymbol{r}) \boldsymbol{r} \phi_{n'\boldsymbol{\kappa}'}(\boldsymbol{r}) d^3r = \delta_{n,n'} \frac{1}{i} \int_\Omega \phi^*_{n\boldsymbol{\kappa}}(\boldsymbol{r}) \nabla_{\boldsymbol{\kappa}'} \phi_{n\boldsymbol{\kappa}'}(\boldsymbol{r}) d^3r + \boldsymbol{X}_{nn'}(\boldsymbol{\kappa}) \delta_{\boldsymbol{\kappa},\boldsymbol{\kappa}'} , \tag{14}$$

where $\boldsymbol{X}_{nn'}(\boldsymbol{\kappa})$ denotes the Berry connection matrix of the magnetic Bloch state, and is defined by

$$\boldsymbol{X}_{nn'}(\boldsymbol{\kappa}) = -\frac{1}{i} \int_\Omega u^*_{n\boldsymbol{\kappa}}(\boldsymbol{r}) \nabla_{\boldsymbol{\kappa}} u_{n'\boldsymbol{\kappa}}(\boldsymbol{r}) d^3r . \tag{15}$$

The derivation of Eq. (14) is given in Appendix A. Note that the diagonal element of the Berry connection matrix $\boldsymbol{X}_{nn}(\boldsymbol{\kappa})$ corresponds to the Berry connection for the magnetic Bloch state with the band index of $n$. Appendix B provides a discussion related to the difficulties associated with the position operator $\boldsymbol{r}$. Substituting Eq. (14) into Eq. (9), we get a general expression of velocity for the magnetic Bloch electron:

$$\begin{aligned} \frac{d}{dt}\langle \boldsymbol{r} \rangle &= \frac{1}{\hbar} \sum_{n\boldsymbol{\kappa}} \sum_{\boldsymbol{\kappa}'} a^*_n(\boldsymbol{\kappa},t) a_n(\boldsymbol{\kappa}',t) \{\varepsilon_n(\boldsymbol{\kappa}) - \varepsilon_n(\boldsymbol{\kappa}')\} \int_\Omega \phi^*_{n\boldsymbol{\kappa}}(\boldsymbol{r}) \nabla_{\boldsymbol{\kappa}'} \phi_{n\boldsymbol{\kappa}'}(\boldsymbol{r}) d^3r \\ &\quad - \frac{1}{i\hbar} \sum_{n\boldsymbol{\kappa}} \sum_{n'} a^*_n(\boldsymbol{\kappa},t) a_{n'}(\boldsymbol{\kappa},t) \{\varepsilon_n(\boldsymbol{\kappa}) - \varepsilon_{n'}(\boldsymbol{\kappa})\} \boldsymbol{X}_{nn'}(\boldsymbol{\kappa}). \end{aligned} \tag{16}$$

In the case of the single-band approximation, substituting Eq. (14) into only the first term of Eq. (11), we get

$$\begin{aligned}\frac{d}{dt}\langle \boldsymbol{r}\rangle_n^{\mathrm{SB}} &= \frac{1}{\hbar}\sum_{n\boldsymbol{\kappa}}\sum_{\boldsymbol{\kappa}'} a^*(\boldsymbol{\kappa},t)a(\boldsymbol{\kappa}',t)\left\{\varepsilon_n(\boldsymbol{\kappa})-\varepsilon_n(\boldsymbol{\kappa}')\right\}\int_\Omega \phi^*_{n\boldsymbol{\kappa}}(\boldsymbol{r})\nabla_{\boldsymbol{\kappa}'}\phi_{n\boldsymbol{\kappa}'}(\boldsymbol{r})d^3r \\ &\quad -\frac{1}{i\hbar}\sum_{\boldsymbol{\kappa}}\sum_{\boldsymbol{\kappa}'} a^*(\boldsymbol{\kappa},t)a(\boldsymbol{\kappa}',t)\sum_{\boldsymbol{\kappa}''}\Big\{\int_\Omega \phi^*_{n\boldsymbol{\kappa}}(\boldsymbol{r}')\boldsymbol{r}'\phi_{n\boldsymbol{\kappa}''}(\boldsymbol{r}')d^3r'\int_\Omega \phi^*_{n\boldsymbol{\kappa}''}(\boldsymbol{r})\left(\boldsymbol{F}\cdot\boldsymbol{r}\right)\phi_{n\boldsymbol{\kappa}'}(\boldsymbol{r})d^3r \\ &\qquad -\int_\Omega \phi^*_{n\boldsymbol{\kappa}}(\boldsymbol{r}')\left(\boldsymbol{F}\cdot\boldsymbol{r}'\right)\phi_{n\boldsymbol{\kappa}''}(\boldsymbol{r}')d^3r'\int_\Omega \phi^*_{n\boldsymbol{\kappa}''}(\boldsymbol{r})\boldsymbol{r}\phi_{n\boldsymbol{\kappa}'}(\boldsymbol{r})d^3r\Big\}. \end{aligned} \tag{17}$$

It should be noted that the second term in Eq. (16) vanishes in Eq. (17) due to the single-band approximation. Since the anomalous velocities perpendicular to the electric field can be obtained from the second term in Eq. (16) when multiple energy bands are correctly considered, and from the second term in Eq. (17) when the single-band approximation is used, the following discussion will focus on these terms. If we denote the second terms in Eq. (16) and (17) as $\boldsymbol{v}_{\mathrm{ano}}$ and $\boldsymbol{v}_{\mathrm{ano}}^{\mathrm{SB},n}$, respectively, then we have

$$\boldsymbol{v}_{\mathrm{ano}} = -\frac{1}{i\hbar}\sum_{n\boldsymbol{\kappa}}\sum_{n'} a_n^*(\boldsymbol{\kappa},t)a_{n'}(\boldsymbol{\kappa},t)\left\{\varepsilon_n(\boldsymbol{\kappa})-\varepsilon_{n'}(\boldsymbol{\kappa})\right\}\boldsymbol{X}_{nn'}(\boldsymbol{\kappa}), \tag{18}$$

$$\begin{aligned}\boldsymbol{v}_{\mathrm{ano}}^{\mathrm{SB},n} &= -\frac{1}{i\hbar}\sum_{\boldsymbol{\kappa}}\sum_{\boldsymbol{\kappa}'} a^*(\boldsymbol{\kappa},t)a(\boldsymbol{\kappa}',t)\sum_{\boldsymbol{\kappa}''}\Big\{\int_\Omega \phi^*_{n\boldsymbol{\kappa}}(\boldsymbol{r}')\boldsymbol{r}'\phi_{n\boldsymbol{\kappa}''}(\boldsymbol{r}')d^3r'\int_\Omega \phi^*_{n\boldsymbol{\kappa}''}(\boldsymbol{r})\left(\boldsymbol{F}\cdot\boldsymbol{r}\right)\phi_{n\boldsymbol{\kappa}'}(\boldsymbol{r})d^3r \\ &\qquad -\int_\Omega \phi^*_{n\boldsymbol{\kappa}}(\boldsymbol{r}')\left(\boldsymbol{F}\cdot\boldsymbol{r}'\right)\phi_{n\boldsymbol{\kappa}''}(\boldsymbol{r}')d^3r'\int_\Omega \phi^*_{n\boldsymbol{\kappa}''}(\boldsymbol{r})\boldsymbol{r}\phi_{n\boldsymbol{\kappa}'}(\boldsymbol{r})d^3r\Big\}. \end{aligned} \tag{19}$$

It is stressed again that while $\boldsymbol{v}_{\mathrm{ano}}$ is derived from Eq. (9), $\boldsymbol{v}_{\mathrm{ano}}^{\mathrm{SB},n}$ is from the second term of Eq. (11) which vanishes when we properly account for the multiple energy bands.

**A. Anomalous velocity in considering multiple energy bands**

First, we discuss the anomalous velocity in considering multiple energy bands $\boldsymbol{v}_{\mathrm{ano}}$. The electric field dependence of $\boldsymbol{v}_{\mathrm{ano}}$ comes from that of the coefficient $a_n(\boldsymbol{\kappa},t)$. In order to discuss the electric field dependence of $a_n(\boldsymbol{\kappa},t)$, we adopt the perturbation theory in which $\hat{H}_0$ and $-\boldsymbol{F}\cdot\boldsymbol{r}$ are treated as the unperturbed Hamiltonian and perturbation term, respectively. Within the first-order perturbation theory, the corrected wave function for the unperturbed state $\phi_{n_0\boldsymbol{\kappa}_0}(\boldsymbol{r})$ is given by

$$\bar{\psi}_{n_0\boldsymbol{\kappa}_0}(\boldsymbol{r},t) = e^{i\frac{\bar{\varepsilon}_{n_0}(\boldsymbol{\kappa}_0)}{\hbar}t}\left\{\phi_{n_0\boldsymbol{\kappa}_0}(\boldsymbol{r}) + \sum_{(n\boldsymbol{\kappa})\neq(n_0\boldsymbol{\kappa}_0)} \frac{\int_\Omega \phi^*_{n\boldsymbol{\kappa}}(\boldsymbol{r})(-\boldsymbol{F}\cdot\boldsymbol{r})\phi_{n_0\boldsymbol{\kappa}_0}(\boldsymbol{r})d^3r}{\varepsilon_{n_0}(\boldsymbol{\kappa}_0)-\varepsilon_n(\boldsymbol{\kappa})}\phi_{n\boldsymbol{\kappa}}(\boldsymbol{r})\right\}, \tag{20}$$

where we assume that there is no degeneracy of the eigenstate $\phi_{n_0\boldsymbol{\kappa}_0}(\boldsymbol{r})$, and where $\bar{\varepsilon}_{n_0}(\boldsymbol{\kappa}_0)$ denotes the corrected eigenvalue. From Eq. (20), we have the approximate form of the coefficient:

$$a_n^{n_0\boldsymbol{\kappa}_0}(\boldsymbol{\kappa},t) = e^{i\frac{\bar{\varepsilon}_{n_0}(\boldsymbol{\kappa}_0)}{\hbar}t}\left\{\delta_{n\boldsymbol{\kappa},n_0\boldsymbol{\kappa}_0} + \sum_{(n\boldsymbol{\kappa})\neq(n_0\boldsymbol{\kappa}_0)} \frac{\int_\Omega \phi^*_{n\boldsymbol{\kappa}}(\boldsymbol{r})(-\boldsymbol{F}\cdot\boldsymbol{r})\phi_{n_0\boldsymbol{\kappa}_0}(\boldsymbol{r})d^3r}{\varepsilon_{n_0}(\boldsymbol{\kappa}_0)-\varepsilon_n(\boldsymbol{\kappa})}\right\}. \tag{21}$$

Substituting Eq. (21) into Eq. (18), we may obtain the expression for the velocity that comes from the magnetic Bloch state $\phi_{n_0\boldsymbol{\kappa}_0}(\boldsymbol{r})$. If we denote the velocity that comes from the magnetic Bloch state $\phi_{n_0\boldsymbol{\kappa}_0}(\boldsymbol{r})$ by $\boldsymbol{v}_{\text{ano}}(n_0,\boldsymbol{\kappa}_0)$, then the velocity term that is proportional to the electric field is given by

$$\boldsymbol{v}_{\text{ano}}(n_0,\boldsymbol{\kappa}_0) \approx \boldsymbol{v}_1(n_0,\boldsymbol{\kappa}_0) + \boldsymbol{v}_2(n_0,\boldsymbol{\kappa}_0) \tag{22}$$

with

$$\boldsymbol{v}_1(n_0,\boldsymbol{\kappa}_0) = \frac{1}{i\hbar}\sum_{\boldsymbol{\kappa}\neq\boldsymbol{\kappa}_0}\boldsymbol{F}\times\left\{\int_\Omega \phi^*_{n_0\boldsymbol{\kappa}}(\boldsymbol{r})\nabla_{\boldsymbol{\kappa}_0}\phi_{n_0\boldsymbol{\kappa}_0}(\boldsymbol{r})d^3r \times \int_\Omega \phi^*_{n_0\boldsymbol{\kappa}_0}(\boldsymbol{r})\nabla_{\boldsymbol{\kappa}}\phi_{n_0\boldsymbol{\kappa}}(\boldsymbol{r})d^3r\right\}, \tag{23}$$

$$\boldsymbol{v}_2(n_0,\boldsymbol{\kappa}_0) = \frac{1}{i\hbar}\sum_n \boldsymbol{F}\times\left\{\boldsymbol{X}_{n_0 n}(\boldsymbol{\kappa}_0)\times\boldsymbol{X}_{n n_0}(\boldsymbol{\kappa}_0)\right\}, \tag{24}$$

where we use Eq. (14) and the vector triple product expansion $\boldsymbol{A}\times(\boldsymbol{B}\times\boldsymbol{C}) = \boldsymbol{B}(\boldsymbol{A}\cdot\boldsymbol{C}) - \boldsymbol{C}(\boldsymbol{A}\cdot\boldsymbol{B})$ in derivations of Eqs. (23) and (24). Equations (23) and (24) suggest that both $\boldsymbol{v}_1(n_0,\boldsymbol{\kappa}_0)$ and $\boldsymbol{v}_2(n_0,\boldsymbol{\kappa}_0)$ are perpendicular to the electric field.

Next step is to evaluate $\boldsymbol{v}_1(n_0,\boldsymbol{\kappa}_0)$ and $\boldsymbol{v}_2(n_0,\boldsymbol{\kappa}_0)$ with considering the singularity of the magnetic Bloch function in the MBZ. As mentioned in Sec. I, the MBZ can be divided into multiple patches, and the phase of the magnetic Bloch function is smooth within each patch, but discontinuous at the boundaries between patches [31]. The schematic diagram of the MBZ is shown in Fig. 1, in which the MBZ can be divided into two patches that are denoted by $V_I$ and $V_{II}$. For later convenience, the phase mismatch of the magnetic Bloch function at the boundary ($S_b$) between two regions is supposed to be given by $\theta(\boldsymbol{\kappa})$. That is to say, $\phi^{II}_{\alpha\boldsymbol{\kappa}}(\boldsymbol{r})/\phi^{I}_{\alpha\boldsymbol{\kappa}}(\boldsymbol{r}) = e^{i\theta(\boldsymbol{\kappa})}$ at the boundary $S_b$, where $\phi^{I}_{\alpha\boldsymbol{\kappa}}(\boldsymbol{r})$ and $\phi^{II}_{\alpha\boldsymbol{\kappa}}(\boldsymbol{r})$ denote the magnetic Bloch function in $V_I$ and $V_{II}$, respectively.

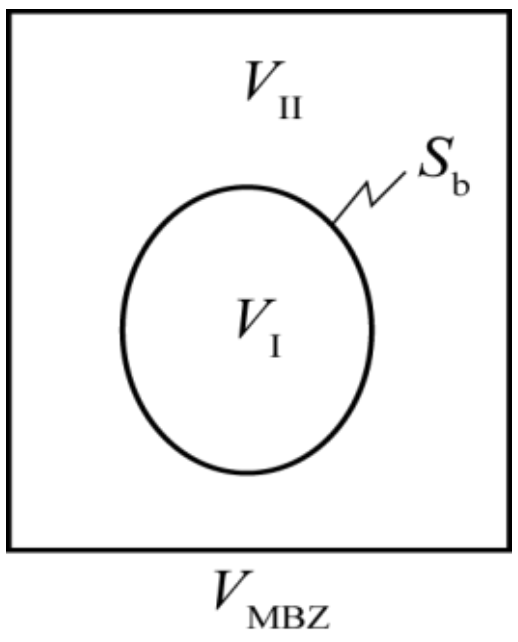

Fig. 1: Schematic diagram of the MBZ. The region of the MBZ ($V_{MBZ}$) is supposed to be divided into two regions $V_I$ and $V_{II}$. The boundary between $V_I$ and $V_{II}$ is denoted as $S_b$.

Based on the discontinuity of the magnetic Bloch function in the MBZ, we evaluate $\boldsymbol{v}_1(n_0,\boldsymbol{\kappa}_0)$ and $\boldsymbol{v}_2(n_0,\boldsymbol{\kappa}_0)$. From Eq. (2), we have

$$\nabla_{\boldsymbol{\kappa}'}\hat{H}_0\phi_{n'\boldsymbol{\kappa}'}(\boldsymbol{r}) = \phi_{n'\boldsymbol{\kappa}'}(\boldsymbol{r})\nabla_{\boldsymbol{\kappa}'}\varepsilon_{n'}(\boldsymbol{\kappa}') + \varepsilon_{n'}(\boldsymbol{\kappa}')\nabla_{\boldsymbol{\kappa}'}\phi_{n'\boldsymbol{\kappa}'}(\boldsymbol{r})\,. \tag{25}$$

In general, if a function is twice differentiable, then the order of partial derivatives is interchangeable [56]. In other words, if a function is not twice differentiable, then the order of partial derivatives is not always interchangeable. Since the magnetic Bloch function is discontinuous in $\boldsymbol{\kappa}$-space at the boundary $S_b$, then the derivative of $\phi_{n'\boldsymbol{\kappa}'}(\boldsymbol{r})$ by $\boldsymbol{r}$ is not always interchangeable with that by $\boldsymbol{\kappa}'$. That is, we have

$$\begin{cases} \nabla_{\boldsymbol{\kappa}'}\hat{H}_0\phi_{n'\boldsymbol{\kappa}'}(\boldsymbol{r}) = \hat{H}_0\nabla_{\boldsymbol{\kappa}'}\phi_{n'\boldsymbol{\kappa}'}(\boldsymbol{r}) & (\boldsymbol{\kappa}' \notin S_b), \\ \nabla_{\boldsymbol{\kappa}'}\hat{H}_0\phi_{n'\boldsymbol{\kappa}'}(\boldsymbol{r}) \neq \hat{H}_0\nabla_{\boldsymbol{\kappa}'}\phi_{n'\boldsymbol{\kappa}'}(\boldsymbol{r}) & (\boldsymbol{\kappa}' \in S_b). \end{cases} \tag{26}$$

Multiplying $\phi^*_{n\boldsymbol{\kappa}}(\boldsymbol{r})$ on both sides of Eq. (25), and using Eq. (26), we get

$$\left\{\varepsilon_n(\boldsymbol{\kappa}) - \varepsilon_{n'}(\boldsymbol{\kappa}')\right\}\int_\Omega \phi^*_{n\boldsymbol{\kappa}}(\boldsymbol{r})\nabla_{\boldsymbol{\kappa}'}\phi_{n'\boldsymbol{\kappa}'}(\boldsymbol{r})d^3r = \delta_{n'\boldsymbol{\kappa}',n\boldsymbol{\kappa}}\nabla_{\boldsymbol{\kappa}'}\varepsilon_{n'}(\boldsymbol{\kappa}') \quad \text{for } \boldsymbol{\kappa}' \notin S_b\,. \tag{27}$$

From Eq. (27) and the assumption of non-degeneracy, we have the following result for $(n\boldsymbol{\kappa}) \neq (n'\boldsymbol{\kappa}')$:

$$\int_\Omega \phi^*_{n\boldsymbol{\kappa}}(\boldsymbol{r})\nabla_{\boldsymbol{\kappa}'}\phi_{n'\boldsymbol{\kappa}'}(\boldsymbol{r})d^3r = \begin{cases} 0 & \text{for } \boldsymbol{\kappa}' \notin S_b \\ \int_\Omega \phi^*_{n\boldsymbol{\kappa}}(\boldsymbol{r})\nabla_{\boldsymbol{\kappa}'}\phi_{n'\boldsymbol{\kappa}'}(\boldsymbol{r})d^3r & \text{for } \boldsymbol{\kappa}' \in S_b \end{cases} \tag{28}$$

Note that in the case of $\boldsymbol{\kappa}' \in S_b$, we can only say that $\int_{\Omega} \phi^*_{n\boldsymbol{\kappa}}(\boldsymbol{r}) \nabla_{\boldsymbol{\kappa}'} \phi_{n'\boldsymbol{\kappa}'}(\boldsymbol{r}) d^3r$ is a vector determined by $(n\boldsymbol{\kappa})$ and $(n'\boldsymbol{\kappa}')$. Using Eq. (28), we have

$$\boldsymbol{v}_1(n_0,\boldsymbol{\kappa}_0) = \begin{cases} 0 & (\boldsymbol{\kappa}_0 \notin S_b) \\ \dfrac{1}{i\hbar} \sum\limits_{\boldsymbol{\kappa} \neq \boldsymbol{\kappa}_0}^{S_b} \boldsymbol{F} \times \left\{ \int_{\Omega} \phi^*_{n_0\boldsymbol{\kappa}}(\boldsymbol{r}) \nabla_{\boldsymbol{\kappa}_0} \phi_{n_0\boldsymbol{\kappa}_0}(\boldsymbol{r}) d^3r \times \int_{\Omega} \phi^*_{n_0\boldsymbol{\kappa}_0}(\boldsymbol{r}) \nabla_{\boldsymbol{\kappa}} \phi_{n_0\boldsymbol{\kappa}}(\boldsymbol{r}) d^3r \right\} & (\boldsymbol{\kappa}_0 \in S_b). \end{cases} \tag{29}$$

Also, $\boldsymbol{v}_2(n_0,\boldsymbol{\kappa}_0)$ is rewritten by

$$\boldsymbol{v}_2(n_0,\boldsymbol{\kappa}_0) = \begin{cases} -\dfrac{1}{\hbar} \boldsymbol{F} \times \Omega_{n_0 n_0}(\boldsymbol{\kappa}_0) & (\boldsymbol{\kappa}_0 \notin S_b) \\ -\dfrac{1}{\hbar} \boldsymbol{F} \times \Omega_{n_0 n_0}(\boldsymbol{\kappa}_0) - \dfrac{1}{i\hbar} \boldsymbol{F} \times \left[ \int_{\Omega} u^*_{n_0\boldsymbol{\kappa}_0}(\boldsymbol{r}) \nabla_{\boldsymbol{\kappa}_0} \times \left\{ \nabla_{\boldsymbol{\kappa}_0} u_{n_0\boldsymbol{\kappa}_0}(\boldsymbol{r}) \right\} d^3r \right] & (\boldsymbol{\kappa}_0 \in S_b) \end{cases} \tag{30}$$

where $\boldsymbol{\Omega}_{n_0 n_0}(\boldsymbol{\kappa}_0)$ denotes the Berry curvature of the magnetic Bloch state and is defined as $\boldsymbol{\Omega}_{n_0 n_0}(\boldsymbol{\kappa}) = \nabla_{\boldsymbol{\kappa}} \times \boldsymbol{X}_{n_0 n_0}(\boldsymbol{\kappa})$. Note that if we assume that the derivative of $u_{n_0\boldsymbol{\kappa}_0}(\boldsymbol{r})$ by $\kappa_{0i}$ ($i = x, y, z$) and the derivative by $\kappa_{0j}$ ($j \neq i$) are interchangeable, then $\nabla_{\boldsymbol{\kappa}_0} \times \left\{ \nabla_{\boldsymbol{\kappa}_0} u_{n_0\boldsymbol{\kappa}_0}(\boldsymbol{r}) \right\}$ in Eq. (30) vanishes. However, that is not always true due to the singularity of the magnetic Bloch function [56]. Therefore, since $\nabla_{\boldsymbol{\kappa}_0} \times \left\{ \nabla_{\boldsymbol{\kappa}_0} u_{n_0\boldsymbol{\kappa}_0}(\boldsymbol{r}) \right\}$ does not always vanish at the boundary $S_b$, due to the singularity of the magnetic Bloch function at the boundary $S_b$, we shall leave the second term of Eq. (30) for $\boldsymbol{\kappa}_0 \in S_b$. Substituting Eqs. (29) and (30) into Eq. (22), we finally get

$$\boldsymbol{v}_{\text{ano}}(n_0,\boldsymbol{\kappa}_0) = \begin{cases} -\dfrac{1}{\hbar} \boldsymbol{F} \times \Omega_{n_0 n_0}(\boldsymbol{\kappa}_0) & (\boldsymbol{\kappa}_0 \notin S_b), \\ -\dfrac{1}{\hbar} \boldsymbol{F} \times \Omega_{n_0 n_0}(\boldsymbol{\kappa}_0) - \dfrac{1}{i\hbar} \boldsymbol{F} \times \left[ \int_{\Omega} u^*_{n_0\boldsymbol{\kappa}_0}(\boldsymbol{r}) \nabla_{\boldsymbol{\kappa}_0} \times \left\{ \nabla_{\boldsymbol{\kappa}_0} u_{n_0\boldsymbol{\kappa}_0}(\boldsymbol{r}) \right\} d^3r \right] & \\ \quad + \dfrac{1}{i\hbar} \sum\limits_{\boldsymbol{\kappa} \neq \boldsymbol{\kappa}_0}^{S_b} \boldsymbol{F} \times \left\{ \int_{\Omega} \phi^*_{n_0\boldsymbol{\kappa}}(\boldsymbol{r}) \nabla_{\boldsymbol{\kappa}_0} \phi_{n_0\boldsymbol{\kappa}_0}(\boldsymbol{r}) d^3r \times \int_{\Omega} \phi^*_{n_0\boldsymbol{\kappa}_0}(\boldsymbol{r}) \nabla_{\boldsymbol{\kappa}} \phi_{n_0\boldsymbol{\kappa}}(\boldsymbol{r}) d^3r \right\} & (\boldsymbol{\kappa}_0 \in S_b). \end{cases} \tag{31}$$

Thus, two types of unconventional anomalous velocity appear at the boundary $S_b$ where the phase of the magnetic Bloch function changes discontinuously. These unconventional anomalous velocities are perpendicular to the electric field similarly to the conventional anomalous velocity. Since the phase discontinuity of the magnetic Bloch function is the cause of the nonzero Chern number, it is reasonable to conclude that the unconventional anomalous velocities arising from the same cause can indeed exist.

### B. Anomalous velocity under the single-band approximation

Even though $\boldsymbol{v}_{\text{ano}}^{\text{SB},n}$ obtained in the single-band approximation should essentially be a vanishing velocity term, we shall consider it to understand the difference between the more accurate treatment mentioned in the previous subsection (Sec. III A) and the single-band approximation. Substituting Eq. (14) into Eq. (19), we get

$$\boldsymbol{v}_{\text{ano}}^{\text{SB},n} = -\frac{1}{\hbar}\sum_{\boldsymbol{\kappa}}\left|a(\boldsymbol{\kappa},t)\right|^2 \boldsymbol{F}\times\boldsymbol{\Omega}_{nn}(\boldsymbol{\kappa}) + \sum_{\boldsymbol{\kappa}}\boldsymbol{v}_{\text{another}}^{\text{SB},n}(\boldsymbol{\kappa}), \tag{32}$$

where the first term corresponds to the conventional anomalous velocity. The second term $\boldsymbol{v}_{\text{another}}^{\text{SB},n}(\boldsymbol{\kappa})$ is given by

$$\begin{aligned}
\boldsymbol{v}_{\text{another}}^{\text{SB},n}(\boldsymbol{\kappa}) = &\frac{1}{i\hbar}a^*(\boldsymbol{\kappa},t)\sum_{\boldsymbol{\kappa}''}\int_\Omega \phi_{n\boldsymbol{\kappa}}^*(\boldsymbol{r})\left\{\nabla_{\boldsymbol{\kappa}''}\phi_{n\boldsymbol{\kappa}''}(\boldsymbol{r})\right\}d^3r\int_\Omega \phi_{n\boldsymbol{\kappa}''}^*(\boldsymbol{r})\boldsymbol{F}\cdot\left[\frac{V}{(2\pi)^3}\int_{S_b}a(\boldsymbol{\kappa}',t)\phi_{n\boldsymbol{\kappa}'}^I(\boldsymbol{r})(1-e^{i\theta(\boldsymbol{\kappa}')})d\boldsymbol{S}_{\boldsymbol{\kappa}'}\right]d^3r\\
&-\frac{1}{i\hbar}a^*(\boldsymbol{\kappa},t)\sum_{\boldsymbol{\kappa}''}\int_\Omega \phi_{n\boldsymbol{\kappa}}^*(\boldsymbol{r})\boldsymbol{F}\cdot\left\{\nabla_{\boldsymbol{\kappa}''}\phi_{n\boldsymbol{\kappa}''}(\boldsymbol{r})\right\}d^3r\int_\Omega \phi_{n\boldsymbol{\kappa}''}^*(\boldsymbol{r})\left[\frac{V}{(2\pi)^3}\int_{S_b}a(\boldsymbol{\kappa}',t)\phi_{n\boldsymbol{\kappa}'}^I(\boldsymbol{r})(1-e^{i\theta(\boldsymbol{\kappa}')})d\boldsymbol{S}_{\boldsymbol{\kappa}'}\right]d^3r\\
&+\frac{1}{i\hbar}a^*(\boldsymbol{\kappa},t)\int_\Omega \phi_{n\boldsymbol{\kappa}}^*(\boldsymbol{r})\left[\frac{V}{(2\pi)^3}\int_{S_b}\phi_{n\boldsymbol{\kappa}'}^I(\boldsymbol{r})(1-e^{i\theta(\boldsymbol{\kappa}')})\left\{\nabla_{\boldsymbol{\kappa}'}a(\boldsymbol{\kappa}',t)\right\}(\boldsymbol{F}\cdot d\boldsymbol{S}_{\boldsymbol{\kappa}'})\right]d^3r\\
&-\frac{1}{i\hbar}a^*(\boldsymbol{\kappa},t)\int_\Omega \phi_{n\boldsymbol{\kappa}}^*(\boldsymbol{r})\left[\frac{V}{(2\pi)^3}\int_{S_b}\phi_{n\boldsymbol{\kappa}'}^I(\boldsymbol{r})(1-e^{i\theta(\boldsymbol{\kappa}')})\left\{(\boldsymbol{F}\cdot\nabla_{\boldsymbol{\kappa}'})a(\boldsymbol{\kappa}',t)\right\}d\boldsymbol{S}_{\boldsymbol{\kappa}'}\right]d^3r\\
&-\frac{1}{\hbar}a^*(\boldsymbol{\kappa},t)\int_\Omega \phi_{n\boldsymbol{\kappa}}^*(\boldsymbol{r})\left[\frac{V}{(2\pi)^3}\int_{S_b}\phi_{n\boldsymbol{\kappa}'}^I(\boldsymbol{r})(1-e^{i\theta(\boldsymbol{\kappa}')})a(\boldsymbol{\kappa}',t)\boldsymbol{X}_\alpha(\boldsymbol{\kappa}')(\boldsymbol{F}\cdot d\boldsymbol{S}_{\boldsymbol{\kappa}'})\right]d^3r\\
&-\frac{1}{\hbar}a^*(\boldsymbol{\kappa},t)\left\{\boldsymbol{F}\cdot\boldsymbol{X}_{nn}(\boldsymbol{\kappa})\right\}\int_\Omega \phi_{n\boldsymbol{\kappa}}^*(\boldsymbol{r})\left[\frac{V}{(2\pi)^3}\int_{S_b}a(\boldsymbol{\kappa}',t)\phi_{n\boldsymbol{\kappa}'}^I(\boldsymbol{r})(1-e^{i\theta(\boldsymbol{\kappa}')})d\boldsymbol{S}_{\boldsymbol{\kappa}'}\right]d^3r\\
&+\frac{1}{\hbar}a^*(\boldsymbol{\kappa},t)\int_\Omega \phi_{n\boldsymbol{\kappa}}^*(\boldsymbol{r})\left[\frac{V}{(2\pi)^3}\int_{S_b}\phi_{n\boldsymbol{\kappa}'}^I(\boldsymbol{r})(1-e^{i\theta(\boldsymbol{\kappa}')})\left\{\boldsymbol{F}\cdot\boldsymbol{X}_{nn}(\boldsymbol{\kappa}')\right\}a(\boldsymbol{\kappa}',t)d\boldsymbol{S}_{\boldsymbol{\kappa}'}\right]d^3r\\
&+\frac{1}{\hbar}a^*(\boldsymbol{\kappa},t)\boldsymbol{X}_n(\boldsymbol{\kappa})\int_\Omega \phi_{n\boldsymbol{\kappa}}^*(\boldsymbol{r})\boldsymbol{F}\cdot\left[\frac{V}{(2\pi)^3}\int_{S_b}a(\boldsymbol{\kappa}',t)\phi_{n\boldsymbol{\kappa}'}^I(\boldsymbol{r})(1-e^{i\theta(\boldsymbol{\kappa}')})d\boldsymbol{S}_{\boldsymbol{\kappa}'}\right]d^3r,
\end{aligned} \tag{33}$$

In the derivation of Eq. (33), we replace the summation with respect to $\boldsymbol{\kappa}'$ by the volume integral with respect to $\boldsymbol{\kappa}'$, and account for the discontinuity of the magnetic Bloch function and the periodicity of the MBZ. The detail derivations of Eqs. (32) and (33) are given in Appendix C and D. Also in Appendix C, $\boldsymbol{v}_{\text{another}}^{\text{SB},n}(\boldsymbol{\kappa})$ is shown to be perpendicular to the electric field similarly to the conventional anomalous velocity.

From Eq. (32), $\boldsymbol{v}_{\text{ano}}^{\text{SB},n}$ consists of the conventional anomalous velocity (the first term of Eq. (32)) and another type of anomalous velocity $\boldsymbol{v}_{\text{another}}^{\text{SB},n}(\boldsymbol{\kappa})$. The latter component appears to correspond to unconventional anomalous velocities given in Eq. (31) because it comes from the phase discontinuity of the magnetic Bloch function. However, the unconventional anomalous velocities given by Eq. (31) are different from $\boldsymbol{v}_{\text{another}}^{\text{SB},n}(\boldsymbol{\kappa})$. Thus, although the single-band approximation appears to lead to both conventional and another type of anomalous velocity, the derivation is flawed, and therefore the resultant $\boldsymbol{v}_{\text{another}}^{\text{SB},n}(\boldsymbol{\kappa})$ is different from $\boldsymbol{v}_{\text{ano}}(n_0,\boldsymbol{\kappa}_0)$ that is derived by a more appropriate method considering multiple energy bands. If the velocity of the magnetic Bloch state is discussed, it is essential to consider both multiple energy bands and the discontinuity of the magnetic Bloch function.

## IV. ALTERNATIVE WAYS TO DERIVE CONVENTIONAL AND UNCONVENTIONAL ANOMALOUS VELOCITIES

In this section, conventional and unconventional anomalous velocities (Eq. (31)) are rederived by methods different from that used in the previous sections, namely by both steady-state and time-dependent perturbation theories. The former method, as in the previous sections, uses the gauge such that the uniform electric field is generated from a scalar potential, and treats it as a perturbation potential. On the other hand, the latter method uses the gauge such that the uniform electric field is generated from the time-dependent vector potential, and treats it as a perturbation potential.

### A. Revisit of conventional and unconventional anomalous velocities via steady-state perturbation theory

In this section, we rederive conventional and unconventional anomalous velocities using a steady-state perturbation theory. Similarly to the previous sections, the uniform electric field is supposed to be generated from a scalar potential $-\boldsymbol{E}\cdot\boldsymbol{r}$. In this case, the Hamiltonian is given by Eq. (3). The corresponding scalar potential $-\boldsymbol{F}\cdot\boldsymbol{r}$ $(=e\boldsymbol{E}\cdot\boldsymbol{r})$ is treated as a perturbation potential. According to the steady-state perturbation theory of the first order, the wave function for the unperturbed state $\phi_{n_0\boldsymbol{\kappa}_0}(\boldsymbol{r})$ is corrected as

$$\bar{\phi}_{n_0\boldsymbol{\kappa}_0}(\boldsymbol{r}) = \phi_{n_0\boldsymbol{\kappa}_0}(\boldsymbol{r}) + \sum_{(n\boldsymbol{\kappa})\neq(n_0\boldsymbol{\kappa}_0)} \frac{\int_\Omega \phi^*_{n\boldsymbol{\kappa}}(\boldsymbol{r})(-\boldsymbol{F}\cdot\boldsymbol{r})\phi_{n_0\boldsymbol{\kappa}_0}(\boldsymbol{r})d^3r}{\varepsilon_{n_0}(\boldsymbol{\kappa}_0)-\varepsilon_n(\boldsymbol{\kappa})}\phi_{n\boldsymbol{\kappa}}(\boldsymbol{r}) \, . \tag{34}$$

In the case where the Hamiltonian is given by Eq. (3), the velocity operator $\hat{\boldsymbol{v}}$ is given by

$$\hat{\boldsymbol{v}} = \frac{\boldsymbol{p}+e\boldsymbol{A}(\boldsymbol{r})}{m} \, . \tag{35}$$

The expectation value of $\hat{\boldsymbol{v}}$ with respect to $\bar{\phi}_{n_0\boldsymbol{\kappa}_0}(\boldsymbol{r},t)$ is

$$\begin{aligned}
\bar{\boldsymbol{v}}(n_0,\boldsymbol{\kappa}_0) &= \frac{1}{i\hbar}\int_\Omega \phi^*_{n_0\boldsymbol{\kappa}_0}(\boldsymbol{r})\left[\boldsymbol{r},\hat{H}_0\right]\phi_{n_0\boldsymbol{\kappa}_0}(\boldsymbol{r})d^3r \\
&+\frac{1}{i\hbar}\sum_{(n\boldsymbol{\kappa})\neq(n_0\boldsymbol{\kappa}_0)} \frac{\int_\Omega \phi^*_{n\boldsymbol{\kappa}}(\boldsymbol{r})(-\boldsymbol{F}\cdot\boldsymbol{r})\phi_{n_0\boldsymbol{\kappa}_0}(\boldsymbol{r})d^3r}{\varepsilon_{n_0}(\boldsymbol{\kappa}_0)-\varepsilon_n(\boldsymbol{\kappa})}\int_\Omega \phi^*_{n_0\boldsymbol{\kappa}_0}(\boldsymbol{r})\left[\boldsymbol{r},\hat{H}_0\right]\phi_{n\boldsymbol{\kappa}}(\boldsymbol{r})d^3r \\
&+\frac{1}{i\hbar}\sum_{(n\boldsymbol{\kappa})\neq(n_0\boldsymbol{\kappa}_0)} \frac{\int_\Omega \phi_{n\boldsymbol{\kappa}}(\boldsymbol{r})(-\boldsymbol{F}\cdot\boldsymbol{r})\phi^*_{n_0\boldsymbol{\kappa}_0}(\boldsymbol{r})d^3r}{\varepsilon_{n_0}(\boldsymbol{\kappa}_0)-\varepsilon_n(\boldsymbol{\kappa})}\int_\Omega \phi^*_{n\boldsymbol{\kappa}}(\boldsymbol{r})\left[\boldsymbol{r},\hat{H}_0\right]\phi_{n_0\boldsymbol{\kappa}_0}(\boldsymbol{r})d^3r,
\end{aligned} \tag{36}$$

where we use $\{\boldsymbol{p}+e\boldsymbol{A}(\boldsymbol{r})\}/m = \left[\boldsymbol{r},\hat{H}_0\right]/i\hbar$. The first term of Eq. (36) is equal to the group velocity for the magnetic Bloch state $\phi_{n_0\boldsymbol{\kappa}_0}(\boldsymbol{r})$. By using Eq. (14), we have

$$\begin{aligned}
\int_\Omega \phi^*_{n\boldsymbol{\kappa}}(\boldsymbol{r})\left[\boldsymbol{r},\hat{H}_0\right]\phi_{n'\boldsymbol{\kappa}'}(\boldsymbol{r})d^3r &= \delta_{n,n'}i\{\varepsilon_n(\boldsymbol{\kappa})-\varepsilon_{n'}(\boldsymbol{\kappa}')\}\int_\Omega \phi^*_{n\boldsymbol{\kappa}}(\boldsymbol{r})\nabla_{\boldsymbol{\kappa}'}\phi_{n\boldsymbol{\kappa}'}(\boldsymbol{r})d^3r \\
&-\delta_{\boldsymbol{\kappa},\boldsymbol{\kappa}'}\{\varepsilon_n(\boldsymbol{\kappa})-\varepsilon_{n'}(\boldsymbol{\kappa}')\}\boldsymbol{X}_{nn'}(\boldsymbol{\kappa}),
\end{aligned} \tag{37}$$

Substituting Eq. (37) into the second and third terms of Eq. (36), we get

$$\begin{aligned}
\bar{\boldsymbol{v}}(n_0,\boldsymbol{\kappa}_0) &= \frac{1}{\hbar}\nabla_{\boldsymbol{\kappa}_0}\varepsilon_{n_0}(\boldsymbol{\kappa}_0) \\
&-\frac{1}{\hbar}\sum_{\boldsymbol{\kappa}\neq\boldsymbol{\kappa}_0}\Big[\Big\{\boldsymbol{F}\cdot\int_\Omega \phi^*_{n_0\boldsymbol{\kappa}}(\boldsymbol{r})\boldsymbol{r}\phi_{n_0\boldsymbol{\kappa}_0}(\boldsymbol{r})d^3r\Big\}\int_\Omega \phi^*_{n_0\boldsymbol{\kappa}_0}(\boldsymbol{r})\nabla_{\boldsymbol{\kappa}}\phi_{n_0\boldsymbol{\kappa}}(\boldsymbol{r})d^3r \\
&\qquad -\Big\{\boldsymbol{F}\cdot\int_\Omega \phi^*_{n_0\boldsymbol{\kappa}_0}(\boldsymbol{r})\boldsymbol{r}\phi_{n_0\boldsymbol{\kappa}}(\boldsymbol{r})d^3r\Big\}\int_\Omega \phi^*_{n_0\boldsymbol{\kappa}}(\boldsymbol{r})\nabla_{\boldsymbol{\kappa}_0}\phi_{n_0\boldsymbol{\kappa}_0}(\boldsymbol{r})d^3r\Big] \\
&+\frac{1}{i\hbar}\sum_{n\neq n_0}\Big[\Big\{\boldsymbol{F}\cdot\int_\Omega \phi^*_{n\boldsymbol{\kappa}_0}(\boldsymbol{r})\boldsymbol{r}\phi_{n_0\boldsymbol{\kappa}_0}(\boldsymbol{r})d^3r\Big\}\boldsymbol{X}_{n_0n}(\boldsymbol{\kappa}_0) \\
&\qquad -\Big\{\boldsymbol{F}\cdot\int_\Omega \phi^*_{n_0\boldsymbol{\kappa}_0}(\boldsymbol{r})\boldsymbol{r}\phi_{n\boldsymbol{\kappa}_0}(\boldsymbol{r})d^3r\Big\}\boldsymbol{X}_{nn_0}(\boldsymbol{\kappa}_0)\Big].
\end{aligned} \tag{38}$$

Finally, the substitution of Eq. (14) into Eq. (38) leads to

$$\begin{aligned}\bar{\boldsymbol{v}}(n_0,\boldsymbol{\kappa}_0) &= \frac{1}{\hbar}\nabla_{\boldsymbol{\kappa}_0}\varepsilon_{n_0}(\boldsymbol{\kappa}_0) \\ &+\frac{1}{i\hbar}\sum_{\boldsymbol{\kappa}\neq\boldsymbol{\kappa}_0}\boldsymbol{F}\times\left\{\int_\Omega \phi^*_{n_0\boldsymbol{\kappa}}(\boldsymbol{r})\nabla_{\boldsymbol{\kappa}_0}\phi_{n_0\boldsymbol{\kappa}_0}(\boldsymbol{r})d^3r\times\int_\Omega \phi^*_{n_0\boldsymbol{\kappa}_0}(\boldsymbol{r})\nabla_{\boldsymbol{\kappa}}\phi_{n_0\boldsymbol{\kappa}}(\boldsymbol{r})d^3r\right\} \\ &+\frac{1}{i\hbar}\sum_n \boldsymbol{F}\times\left\{\boldsymbol{X}_{n_0 n}(\boldsymbol{\kappa}_0)\times\boldsymbol{X}_{n n_0}(\boldsymbol{\kappa}_0)\right\},\end{aligned} \tag{39}$$

where we use the vector triple product expansion $\boldsymbol{A}\times(\boldsymbol{B}\times\boldsymbol{C})=\boldsymbol{B}(\boldsymbol{A}\cdot\boldsymbol{C})-\boldsymbol{C}(\boldsymbol{A}\cdot\boldsymbol{B})$. According to the definitions of $\boldsymbol{v}_1(n_0,\boldsymbol{\kappa}_0)$ and $\boldsymbol{v}_2(n_0,\boldsymbol{\kappa}_0)$ given in Eqs. (22) and (23), the second and third terms of Eq. (39) are just equal to $\boldsymbol{v}_1(n_0,\boldsymbol{\kappa}_0)$ and $\boldsymbol{v}_2(n_0,\boldsymbol{\kappa}_0)$, respectively. Therefore, as shown in III A, the second and third terms of Eq. (39) is eventually summarized as in Eq. (31). Thus, the conventional and unconventional anomalous velocities can be rederived also using a steady-state perturbation theory.

**B. Revisit of conventional and unconventional anomalous velocities via time-dependent perturbation theory**

In this subsection, we confirm the existence of $\boldsymbol{v}_1(n_0,\boldsymbol{\kappa}_0)$ and $\boldsymbol{v}_2(n_0,\boldsymbol{\kappa}_0)$ by using a time-dependent vector potential yielding a uniform electric field instead of using a scalar potential $-\boldsymbol{F}\cdot\boldsymbol{r}$ $(=e\boldsymbol{E}\cdot\boldsymbol{r})$. A time-dependent vector potential yielding a uniform electric field is simply given by $-\boldsymbol{E}t$. Since the vector potential $-\boldsymbol{E}t$ represents the uniform electric field $\boldsymbol{E}$ that is applied stepwise at $t=0$, it has an unphysical aspect [57]. In actual systems the electric field is gradually applied in the remote past ($t=-\infty$), and becomes an objective one at $t=0$, and then gradually approaches to zero in the remote future ($t=\infty$). The wavefunction becomes a magnetic Bloch state at $t=\pm\infty$. In order to avoid the unphysical effect of a sudden change in the Hamiltonian, we turn on the vector potential gradually to its full value:

$$\boldsymbol{A}_\alpha(t)=-\boldsymbol{E}te^{-\alpha|t|}, \tag{40}$$

where $e^{-\alpha|t|}$ is an adiabatic factor and $\alpha$ is small and positive [57,58]. Note that $-\partial\boldsymbol{A}_\alpha(t)/\partial t\approx\boldsymbol{E}$ for $\alpha|t|\ll 1$ and that the electric field is fully applied at $t=0$. At the end of calculations, we let $\alpha\to 0$ and evaluate the expectation value of the velocity operator at $t=0$.

The Hamiltonian for an electron under both the periodic potential $V(\boldsymbol{r})$ of a crystal and the vector potential $\boldsymbol{A}(\boldsymbol{r})+\boldsymbol{A}_\alpha(t)$ is given by

$$\begin{aligned}\hat{H}_\alpha &= \frac{1}{2m}\left\{\boldsymbol{p}+e\boldsymbol{A}(\boldsymbol{r})-e\boldsymbol{E}te^{-\alpha|t|}\right\}^2+V(\boldsymbol{r}) \\ &= \hat{H}_0-\frac{e}{m}te^{-\alpha|t|}\boldsymbol{E}\cdot\left\{\boldsymbol{p}+e\boldsymbol{A}(\boldsymbol{r})\right\}+\frac{e^2}{2m}t^2e^{-2\alpha|t|}\boldsymbol{E}^2.\end{aligned} \tag{41}$$

If we consider the first-order with respect to $\boldsymbol{E}$ , the time-dependent perturbation is given by

$$\hat{H}'_\alpha(t)=-\frac{e}{m}te^{-\alpha|t|}\boldsymbol{E}\cdot\left\{\boldsymbol{p}+e\boldsymbol{A}(\boldsymbol{r})\right\}. \tag{42}$$

The time-dependent Schrödinger equation is given by

$$i\hbar\frac{\partial\psi_\alpha(\boldsymbol{r},t)}{\partial t}=\hat{H}_\alpha\psi_\alpha(\boldsymbol{r},t) \tag{43}$$

The time-dependent wavefunction $\psi_\alpha(\boldsymbol{r},t)$ is expanded by magnetic Bloch functions, i.e.,

$$\psi_\alpha(\boldsymbol{r},t)=\sum_{n\boldsymbol{\kappa}}c_{n\boldsymbol{\kappa},\alpha}^{n_0\boldsymbol{\kappa}_0}(t)e^{-i\varepsilon_n(\boldsymbol{\kappa})t/\hbar}\phi_{n\boldsymbol{\kappa}}(\boldsymbol{r}), \tag{44}$$

where $c_{n\boldsymbol{\kappa},\alpha}^{n_0\boldsymbol{\kappa}_0}(t)$ denotes the expansion coefficient and satisfies the following initial condition:

$$c_{n\boldsymbol{\kappa},\alpha}^{n_0\boldsymbol{\kappa}_0}(-\infty)=\delta_{n\boldsymbol{\kappa},n_0\boldsymbol{\kappa}_0}. \tag{45}$$

The time-dependent perturbation theory provides the approximate form of $c_{n\boldsymbol{\kappa},\alpha}^{n_0\boldsymbol{\kappa}_0}(t)$ within the first order of $\boldsymbol{E}$ :

$$c_{n\boldsymbol{\kappa},\alpha}^{n_0\boldsymbol{\kappa}_0}(t) = \delta_{n\boldsymbol{\kappa},n_0\boldsymbol{\kappa}_0} - \frac{e}{i\hbar m}\int_{-\infty}^{t} t' e^{\left\{i\frac{\varepsilon_n(\boldsymbol{\kappa})-\varepsilon_{n_0}(\boldsymbol{\kappa}_0)}{\hbar}t'-\alpha|t'|\right\}} dt' \boldsymbol{E}\cdot\int_\Omega \phi_{n\boldsymbol{\kappa}}^*(\boldsymbol{r})\left\{\boldsymbol{p}+e\boldsymbol{A}(\boldsymbol{r})\right\}\phi_{n_0\boldsymbol{\kappa}_0}(\boldsymbol{r})d^3r$$

$$= \delta_{n\boldsymbol{\kappa},n_0\boldsymbol{\kappa}_0} - \frac{e}{i\hbar m}\left\{B_\alpha^t(n\boldsymbol{\kappa},n_0\boldsymbol{\kappa}_0)e^{-i\frac{\varepsilon_n(\boldsymbol{\kappa})-\varepsilon_{n_0}(\boldsymbol{\kappa}_0)}{\hbar}t} + B_\alpha^0(n\boldsymbol{\kappa},n_0\boldsymbol{\kappa}_0)\right\} \tag{46}$$

$$\times\boldsymbol{E}\cdot\int_\Omega \phi_{n\boldsymbol{\kappa}}^*(\boldsymbol{r})\left\{\boldsymbol{p}+e\boldsymbol{A}(\boldsymbol{r})\right\}\phi_{n_0\boldsymbol{\kappa}_0}(\boldsymbol{r})d^3r$$

with

$$B_\alpha^t(n\boldsymbol{\kappa},n_0\boldsymbol{\kappa}_0) = \frac{\hbar}{i}\left\{t - \frac{\hbar}{i}\frac{1}{\varepsilon_n(\boldsymbol{\kappa})-\varepsilon_{n_0}(\boldsymbol{\kappa}_0)+i\hbar\alpha}\right\}\frac{e^{-\alpha t}}{\varepsilon_n(\boldsymbol{\kappa})-\varepsilon_{n_0}(\boldsymbol{\kappa}_0)+i\hbar\alpha},$$

$$B_\alpha^0(n\boldsymbol{\kappa},n_0\boldsymbol{\kappa}_0) = \left\{\frac{\hbar}{\varepsilon_n(\boldsymbol{\kappa})-\varepsilon_{n_0}(\boldsymbol{\kappa}_0)-i\hbar\alpha}\right\}^2 - \left\{\frac{\hbar}{\varepsilon_n(\boldsymbol{\kappa})-\varepsilon_{n_0}(\boldsymbol{\kappa}_0)+i\hbar\alpha}\right\}^2. \tag{47}$$

The velocity operator for the Hamiltonian in Eq. (41) is given by

$$\hat{\boldsymbol{v}}' = \frac{\boldsymbol{p}+e\boldsymbol{A}(\boldsymbol{r})+e\boldsymbol{A}_\alpha(t)}{m}. \tag{48}$$

The expectation value of $\hat{\boldsymbol{v}}'$ with respect to $\psi_\alpha(\boldsymbol{r},t)$ is given by

$$\boldsymbol{v}'_\alpha(n_0\boldsymbol{\kappa}_0,t) = \frac{1}{i\hbar}\sum_{n\boldsymbol{\kappa}}\sum_{n'\boldsymbol{\kappa}'} c_{n\boldsymbol{\kappa},\alpha}^{n_0\boldsymbol{\kappa}_0}(t)^* c_{n'\boldsymbol{\kappa}',\alpha}^{n_0\boldsymbol{\kappa}_0}(t) e^{\left\{i\frac{\varepsilon_n(\boldsymbol{\kappa})-\varepsilon_{n'}(\boldsymbol{\kappa}')}{\hbar}t'\right\}}\int_\Omega \phi_{n\boldsymbol{\kappa}}^*(\boldsymbol{r})\left[\boldsymbol{r},\hat{H}_0\right]\phi_{n'\boldsymbol{\kappa}'}(\boldsymbol{r})d^3r - \frac{e\boldsymbol{E}}{m}te^{-\alpha|t|}, \tag{49}$$

where we use $\left\{\boldsymbol{p}+e\boldsymbol{A}(\boldsymbol{r})\right\}/m = \left[\boldsymbol{r},\hat{H}_0\right]/i\hbar$. The substitution of Eq. (37) into Eq. (49) leads to

$$\boldsymbol{v}'_\alpha(n_0\boldsymbol{\kappa}_0,t) = \sum_{n\boldsymbol{\kappa}}\left|c_{n\boldsymbol{\kappa},\alpha}^{n_0\boldsymbol{\kappa}_0}(t)\right|^2\frac{1}{\hbar}\nabla_{\boldsymbol{\kappa}_0}\varepsilon_{n_0}(\boldsymbol{\kappa}_0) - \frac{e\boldsymbol{E}}{m}te^{-\alpha|t|}$$

$$+\frac{1}{\hbar}\sum_{n\boldsymbol{\kappa}}\sum_{\boldsymbol{\kappa}'\neq\boldsymbol{\kappa}} c_{n\boldsymbol{\kappa},\alpha}^{n_0\boldsymbol{\kappa}_0}(t)^* c_{n\boldsymbol{\kappa}',\alpha}^{n_0\boldsymbol{\kappa}_0}(t)e^{\left\{i\frac{\varepsilon_n(\boldsymbol{\kappa})-\varepsilon_n(\boldsymbol{\kappa}')}{\hbar}t\right\}}\{\varepsilon_n(\boldsymbol{\kappa})-\varepsilon_n(\boldsymbol{\kappa}')\}\int_\Omega \phi_{n\boldsymbol{\kappa}}^*(\boldsymbol{r})\nabla_{\boldsymbol{\kappa}'}\phi_{n\boldsymbol{\kappa}'}(\boldsymbol{r})d^3r$$

$$-\frac{1}{i\hbar}\sum_{n\boldsymbol{\kappa}}\sum_{n'\neq n} c_{n\boldsymbol{\kappa},\alpha}^{n_0\boldsymbol{\kappa}_0}(t)^* c_{n'\boldsymbol{\kappa},\alpha}^{n_0\boldsymbol{\kappa}_0}(t)e^{\left\{i\frac{\varepsilon_n(\boldsymbol{\kappa})-\varepsilon_{n'}(\boldsymbol{\kappa})}{\hbar}t\right\}}\{\varepsilon_n(\boldsymbol{\kappa})-\varepsilon_{n'}(\boldsymbol{\kappa})\}\boldsymbol{X}_{nn'}(\boldsymbol{\kappa}). \tag{50}$$

Substituting Eq. (46) into Eq. (50), and neglecting the second-order terms of $\boldsymbol{E}$ , we have

$$
\begin{aligned}
\boldsymbol{v}'_\alpha(n_0\boldsymbol{\kappa}_0,t) &= \sum_{n\boldsymbol{\kappa}} \left| c^{n_0\boldsymbol{\kappa}_0}_{n\boldsymbol{\kappa},\alpha}(t) \right|^2 \frac{1}{\hbar}\nabla_{\boldsymbol{\kappa}_0}\varepsilon_{n_0}(\boldsymbol{\kappa}_0) - \frac{e\boldsymbol{E}}{m} t e^{-\alpha|t|} \\
&+ \frac{e}{i\hbar^3}\sum_{\boldsymbol{\kappa}'\neq\boldsymbol{\kappa}} D_\alpha(n_0\boldsymbol{\kappa},n_0\boldsymbol{\kappa}_0,t)\left\{\boldsymbol{E}\cdot\int_\Omega \phi^*_{n_0\boldsymbol{\kappa}}(\boldsymbol{r})\nabla_{\boldsymbol{\kappa}_0}\phi_{n_0\boldsymbol{\kappa}_0}(\boldsymbol{r})d^3r\right\}\int_\Omega \phi^*_{n_0\boldsymbol{\kappa}_0}(\boldsymbol{r})\nabla_{\boldsymbol{\kappa}}\phi_{n_0\boldsymbol{\kappa}}(\boldsymbol{r})d^3r \\
&- \frac{e}{i\hbar^3}\sum_{\boldsymbol{\kappa}'\neq\boldsymbol{\kappa}} D^*_\alpha(n_0\boldsymbol{\kappa},n_0\boldsymbol{\kappa}_0,t)\left\{\boldsymbol{E}\cdot\int_\Omega \phi^*_{n_0\boldsymbol{\kappa}_0}(\boldsymbol{r})\nabla_{\boldsymbol{\kappa}}\phi_{n_0\boldsymbol{\kappa}}(\boldsymbol{r})d^3r\right\}\int_\Omega \phi^*_{n_0\boldsymbol{\kappa}}(\boldsymbol{r})\nabla_{\boldsymbol{\kappa}_0}\phi_{n_0\boldsymbol{\kappa}_0}(\boldsymbol{r})d^3r \quad (51) \\
&- \frac{e}{i\hbar^3}\sum_{\boldsymbol{\kappa}'\neq\boldsymbol{\kappa}} D_\alpha(n\boldsymbol{\kappa}_0,n_0\boldsymbol{\kappa}_0,t)\left\{\boldsymbol{E}\cdot\boldsymbol{X}_{nn_0}(\boldsymbol{\kappa}_0)\right\}\boldsymbol{X}_{n_0n}(\boldsymbol{\kappa}_0) \\
&- \frac{e}{i\hbar^3}\sum_{\boldsymbol{\kappa}'\neq\boldsymbol{\kappa}} D^*_\alpha(n\boldsymbol{\kappa}_0,n_0\boldsymbol{\kappa}_0,t)\left\{\boldsymbol{E}\cdot\boldsymbol{X}_{n_0n}(\boldsymbol{\kappa}_0)\right\}\boldsymbol{X}_{nn_0}(\boldsymbol{\kappa}_0)
\end{aligned}
$$

with

$$
D_\alpha(n\boldsymbol{\kappa},n_0\boldsymbol{\kappa}_0,t) = \left\{\varepsilon_n(\boldsymbol{\kappa}) - \varepsilon_{n_0}(\boldsymbol{\kappa}_0)\right\}^2 \left\{ B^t_\alpha(n\boldsymbol{\kappa},n_0\boldsymbol{\kappa}_0) + B^0_\alpha(n\boldsymbol{\kappa},n_0\boldsymbol{\kappa}_0) e^{\left\{ i\frac{\varepsilon_{n_0}(\boldsymbol{\kappa}_0)-\varepsilon_n(\boldsymbol{\kappa})}{\hbar} t \right\}} \right\}. \quad (52)
$$

By using Eqs. (47) and (52), we have

$$
\lim_{\alpha\to 0} D_\alpha(n\boldsymbol{\kappa},n_0\boldsymbol{\kappa}_0,t) = \frac{\hbar}{i}\left\{\varepsilon_n(\boldsymbol{\kappa}) - \varepsilon_{n_0}(\boldsymbol{\kappa}_0)\right\} t + \hbar^2 . \quad (53)
$$

Substituting Eq. (53) and $t = 0$ into Eq. (51), we finally get

$$
\begin{aligned}
\boldsymbol{v}'_\alpha(n_0\boldsymbol{\kappa}_0,0) &= \frac{1}{\hbar}\nabla_{\boldsymbol{\kappa}_0}\varepsilon_{n_0}(\boldsymbol{\kappa}_0) + \frac{1}{i\hbar}\sum_{\boldsymbol{\kappa}\neq\boldsymbol{\kappa}_0} \boldsymbol{F}\times\left\{\int_\Omega \phi^*_{n_0\boldsymbol{\kappa}}(\boldsymbol{r})\nabla_{\boldsymbol{\kappa}_0}\phi_{n_0\boldsymbol{\kappa}_0}(\boldsymbol{r})d^3r \times \int_\Omega \phi^*_{n_0\boldsymbol{\kappa}_0}(\boldsymbol{r})\nabla_{\boldsymbol{\kappa}}\phi_{n_0\boldsymbol{\kappa}}(\boldsymbol{r})d^3r\right\} \\
&\frac{1}{i\hbar}\sum_n \boldsymbol{F}\times\left\{\boldsymbol{X}_{n_0n}(\boldsymbol{\kappa}_0)\times\boldsymbol{X}_{nn_0}(\boldsymbol{\kappa}_0)\right\}.
\end{aligned}
\quad (54)
$$

The second and third terms of Eq. (54) just coincides with $\boldsymbol{v}_1(n_0,\boldsymbol{\kappa}_0)$ and $\boldsymbol{v}_2(n_0,\boldsymbol{\kappa}_0)$ , respectively. Thus, the conventional and unconventional anomalous velocities can be rederived also using the time-dependent perturbation theory similarly to the steady-state perturbation theory.

### C. GUAGE INVARIANCE

In the previous Secs. IV A and IV B, the uniform electric field is generated from the scalar potential $-\boldsymbol{E}\cdot\boldsymbol{r}$ and from the vector potential $-\boldsymbol{E}te^{-\alpha|t|}$ with $\alpha\to 0$, respectively. In what follows, the vector potential $-\boldsymbol{E}te^{-\alpha|t|}$ with $\alpha\to 0$ is simply denote by $-\boldsymbol{E}t$. For convenient, the former and latter choices are referred to as gauge 1 and gauge 2, respectively. The gauge transformation from gauge 1 to gauge 2 is given by

$$\begin{aligned}&\boldsymbol{A}(\boldsymbol{r})-\boldsymbol{E}t=\boldsymbol{A}(\boldsymbol{r})+\nabla\chi(\boldsymbol{r},t),\\&-\frac{1}{e}V(\boldsymbol{r})=-\frac{1}{e}V(\boldsymbol{r})-\boldsymbol{E}\cdot\boldsymbol{r}-\frac{\partial\chi(\boldsymbol{r},t)}{\partial t},\end{aligned}\tag{55}$$

where $\chi(\boldsymbol{r},t)$ denotes the gauge function and is given by

$$\chi(\boldsymbol{r},t)=-(\boldsymbol{E}\cdot\boldsymbol{r})t.\tag{56}$$

If the wavefunctions in the gauge 1 and 2 are denoted by $\psi^1(\boldsymbol{r},t)$ and $\psi^2(\boldsymbol{r},t)$, respectively, then we have

$$\psi^2(\boldsymbol{r},t)=e^{i\frac{e}{\hbar}\boldsymbol{E}\cdot\boldsymbol{r}t}\psi^1(\boldsymbol{r},t).\tag{57}$$

Note that in the previous Secs. IV A and IV B, the approximate wavefunctions for $\psi^1(\boldsymbol{r},t)$ and $\psi^2(\boldsymbol{r},t)$ are treated by the steady-state and time-dependent perturbation theories, respectively. Using Eq. (48) with $\alpha\to 0$ and Eq. (57), the expectation value of the velocity operator in the gauge 2 is calculated as

$$\begin{aligned}&\int_\Omega\psi^2(\boldsymbol{r},t)^*\frac{1}{m}\{\boldsymbol{p}+e\boldsymbol{A}(\boldsymbol{r})-e\boldsymbol{E}t\}\psi^2(\boldsymbol{r},t)d^3r\\&=\int_\Omega e^{-i\frac{e}{\hbar}\boldsymbol{E}\cdot\boldsymbol{r}t}\psi^1(\boldsymbol{r},t)^*\frac{1}{m}\{\boldsymbol{p}+e\boldsymbol{A}(\boldsymbol{r})-e\boldsymbol{E}t\}e^{i\frac{e}{\hbar}\boldsymbol{E}\cdot\boldsymbol{r}t}\psi^1(\boldsymbol{r},t)d^3r\\&=\int_\Omega\psi^1(\boldsymbol{r},t)^*\frac{1}{m}\{\boldsymbol{p}+e\boldsymbol{A}(\boldsymbol{r})\}\psi^1(\boldsymbol{r},t)d^3r.\end{aligned}\tag{58}$$

Considering that the velocity operator in gauge 1 is given by Eq. (35), Eq. (58) implies that the same result is obtained whether the velocity of the magnetic Bloch state is estimated in gauge 1 or gauge 2.

The above discussion shows that the velocity is gauge-invariant, while the results of the previous Secs. IV A and IV B show that the velocity calculated using the approximate wave function within the first-order perturbation theories is also gauge-invariant. This suggests that both conventional and unconventional anomalous velocities exist irrespective of the choice of potential describing the uniform electric field.

## V. CONCLUDING REMARKS

We have investigated the velocity of the magnetic Bloch state under a uniform electric field by accounting for both the multiple energy bands and the discontinuity of the magnetic Bloch function in MBZ. The conventional anomalous velocity is found to be actually derived from the term that improperly becomes zero under the single-band approximation. It is also found that in the single-band approximation, the conventional anomalous velocity is derived by accident from the term that vanishes when multiple energy bands are considered. Consequently, the origin of the conventional anomalous velocity under the single-band approximation is incorrect, and the correct origin of the conventional anomalous velocity is a term that becomes nonzero only if the multiple energy bands are considered.

In addition to the conventional anomalous velocity, it is also shown that unconventional types of anomalous velocity may appear due to the phase discontinuity of the magnetic Bloch function in the MBZ. The unconventional anomalous velocities are perpendicular to the electric field similarly to the conventional anomalous velocity. Although another type of anomalous velocity is derived also under the single-band approximation, it comes from the vanishing term, and the expression is different from that of the correct unconventional anomalous velocities that are obtained by considering multiple energy bands appropriately. Additional velocity terms other than the conventional anomalous velocity have also been derived and discussed so far in the field of mathematical physics for solid-state phenomena, but the derivation is based on the single-band approximation [44-47]. Therefore, it is interesting to discuss and revise such the additional velocity terms by considering the multiple energy bands.

Furthermore, it is found that both conventional and unconventional anomalous velocities can also be obtained, regardless of whether the uniform electric field is expressed as the gradient of a scalar potential or derived from the time dependence of a vector potential. This suggests the validity of the existence of both conventional and unconventional anomalous velocities.

As mentioned below Eq. (1), we do not consider the electron spin as in the previous work

[12]. Using the Dirac equation instead of the Schrödinger equation, the effects of the electron spin such as the spin Zeeman effect and spin-orbit interaction can be taken into consideration. This would be one of the future works.

In addition to the above, although it is pointed out that there is a delicate interplay between surface and bulk contributions [55], only the bulk contribution to the velocity is considered in this work. Moreover, the disorder potential, which is one of the main ingredients in describing the integer quantum Hall effect, is not considered in this study. These are also the remaining problems to be addressed in the future.

To discuss how large unconventional anomalous velocities are, we need to calculate $\phi_{n\boldsymbol{\kappa}}(\boldsymbol{r})$ and $\varepsilon_n(\boldsymbol{\kappa})$. For this aim, we may utilize the MFRTB method [24-29]. Numerical calculations would be useful to understand the multiband effects and the physical mechanism of the emergence of unconventional anomalous velocities. In addition, specific numerical calculations would be helpful for devising the detection method for unconventional anomalous velocities. A next challenge would be to apply the MFRTB method to a model system, and then to evaluate unconventional anomalous velocities.

Finally, we give a brief comment on the effect of the electron-electron interaction. It is well-known that the electron-electron interaction plays an important role on the fractional quantum Hall effect [59-62]. In the present formulation of the velocity, the electron-electron interaction is not taken into consideration. For the investigation of the effect of the electron-electron interaction on the velocity, we may employ the density functional theory (DFT) [63,64] and its extensions [65-73]. In such theories, we consider a so-called Kohn-Sham (KS) equation that contains the effect of the electron-electron interaction instead of Eq. (2). For instance, the current-density functional theory (CDFT) [69-73] would be applicable to the system. Suppose that the approximate form of the local density approximation (LDA) functional [64] or generalized gradient approximation (GGA) functional [74] for the DFT is adopted as an approximate form of the exchange-correlation energy functional of the CDFT. Under this approximation, the resultant KS equation of the CDFT would have a form such that $V(\boldsymbol{r})$ in Eq. (3) is replaced by the effective scalar potential of the LDA or GGA. Therefore, the discussion similar to this study may be available. It would be challenging to analyze the effects of the electron-electron interaction on the unconventional anomalous velocities, and furthermore on the fractional quantum Hall effect. The present work will provide a good starting point of such a further study.

**ACKNOWLEDGEMENT**

This work was partially supported by Grant-in-Aid for Scientific Research (No. 18K03510 and No. 18K03461 and No. 23K03250) of the Japan Society for the Promotion of Science. This work was also supported by MEXT KAKENHI Grant Number 24H01618.

**APPENDIX A: Derivation of Eq. (14)**

By using the relation $\nabla_{\kappa'} e^{i\kappa'\cdot r} = i r e^{i\kappa'\cdot r}$, the left-hand side of Eq. (14) can be rewritten by

$$\begin{aligned}\int_\Omega \phi^*_{n\kappa}(\boldsymbol{r})\boldsymbol{r}\phi_{n'\kappa'}(\boldsymbol{r})d^3r &= \frac{1}{i}\int_\Omega \phi^*_{n\kappa}(\boldsymbol{r})\left\{\nabla_{\kappa'} e^{i\kappa'\cdot \boldsymbol{r}}\right\}u_{n'\kappa'}(\boldsymbol{r})d^3r \\ &= \delta_{n,n'}\frac{1}{i}\int_\Omega \phi^*_{n\kappa}(\boldsymbol{r})\nabla_{\kappa'}\phi_{n\kappa'}(\boldsymbol{r})d^3r - \frac{1}{i}\int_\Omega e^{i(\kappa'-\kappa)\cdot \boldsymbol{r}}u^*_{n\kappa}(\boldsymbol{r})\nabla_{\kappa'}u_{n'\kappa'}(\boldsymbol{r})d^3r.\end{aligned} \tag{A1}$$

The integral in the second term of Eq. (A1) can be rewritten by changing the volume integral within the system $\Omega$ to the summation of the volume integral within the magnetic unit cell $\Omega_{unit}$;

$$\int_\Omega e^{i(\kappa'-\kappa)\cdot \boldsymbol{r}}u^*_{n\kappa}(\boldsymbol{r})\nabla_{\kappa}u_{n'\kappa'}(\boldsymbol{r})d^3r = \sum_{\boldsymbol{t}_\alpha} e^{i(\kappa'-\kappa)\cdot \boldsymbol{t}_\alpha}\int_{\Omega_{\text{unit}}} e^{i(\kappa'-\kappa)\cdot \boldsymbol{r}'}u^*_{n\kappa}(\boldsymbol{r}')\nabla_{\kappa}u_{n'\kappa'}(\boldsymbol{r}')d^3r', \tag{A2}$$

where we used Eq. (13). Supposing that the total number of the magnetic primitive unit cells contained in the system is denoted by $N_{\boldsymbol{t}_\alpha}$, it can be shown that

$$\begin{aligned}\sum_{\boldsymbol{t}_\alpha} e^{i(\kappa'-\kappa)\cdot \boldsymbol{t}_\alpha} &= N_{\boldsymbol{t}_\alpha}\delta_{\kappa,\,\kappa'+\boldsymbol{K}} \\ &= N_{\boldsymbol{t}_\alpha}\delta_{\kappa,\,\kappa'},\end{aligned} \tag{A3}$$

where $\boldsymbol{K}$ denotes the magnetic reciprocal lattice vector. Note that since wavevectors $\kappa$ and $\kappa'$ are in the MBZ, $\boldsymbol{K}$ should be equal to zero if $\delta_{\kappa,\,\kappa'+\boldsymbol{K}} = 1$. Substituting Eq. (A3) into the right-hand side of Eq. (A2), we have

$$\int_\Omega e^{i(\kappa'-\kappa)\cdot \boldsymbol{r}}u^*_{n\kappa}(\boldsymbol{r})\nabla_{\kappa'}u_{n'\kappa'}(\boldsymbol{r})d^3r = N_{\boldsymbol{t}_\alpha}\delta_{\kappa,\,\kappa'}\int_{\Omega_{\text{unit}}} u^*_{n\kappa}(\boldsymbol{r}')\nabla_{\kappa'}u_{n'\kappa'}(\boldsymbol{r}')d^3r'. \tag{A4}$$

By using (13), the right-hand side of (A4) can be transformed to the volume integral within $\Omega$, i.e., we have

$$\int_{\Omega} e^{i(\boldsymbol{\kappa}'-\boldsymbol{\kappa})\cdot\boldsymbol{r}} u^*_{n\boldsymbol{\kappa}}(\boldsymbol{r}) \nabla_{\boldsymbol{\kappa}'} u_{n'\boldsymbol{\kappa}'}(\boldsymbol{r}) d^3r = \delta_{\boldsymbol{\kappa},\boldsymbol{\kappa}'} \int_{\Omega} u^*_{n\boldsymbol{\kappa}}(\boldsymbol{r}) \nabla_{\boldsymbol{\kappa}} u_{n'\boldsymbol{\kappa}}(\boldsymbol{r}) d^3r$$
$$= \delta_{\boldsymbol{\kappa},\boldsymbol{\kappa}'} \frac{1}{i} \boldsymbol{X}_{nn'}(\boldsymbol{\kappa}). \tag{A5}$$

Substituting Eq. (A5) into Eq. (A1), the matrix element of $\boldsymbol{r}$ with respect to magnetic Bloch functions can be rewritten as

$$\int_{\Omega} \phi^*_{n\boldsymbol{\kappa}}(\boldsymbol{r}) \boldsymbol{r} \phi_{n'\boldsymbol{\kappa}'}(\boldsymbol{r}) d^3r = \delta_{n,n'} \frac{1}{i} \int_{\Omega} \phi^*_{n\boldsymbol{\kappa}}(\boldsymbol{r}) \nabla_{\boldsymbol{\kappa}'} \phi_{n\boldsymbol{\kappa}'}(\boldsymbol{r}) d^3r + \boldsymbol{X}_{nn'}(\boldsymbol{\kappa}) \delta_{\boldsymbol{\kappa},\boldsymbol{\kappa}'}. \tag{14}$$

**APPENDIX B. Discussion related to the difficulties associated with the position operator**

In general, the position operator $\boldsymbol{r}$, which is included in the scalar potential $V_{applied}(\boldsymbol{r}) = -\boldsymbol{F} \cdot \boldsymbol{r}$, may cause two difficulties. (i) One is that $\boldsymbol{r}$ may go to infinity, as described in the textbook by Kittel [48]. (ii) The other is that the position operator $\boldsymbol{r}$ changes depending on the choice of the coordinate origin. These difficulties can be avoided by following previous works [2-4, 7, 12, 49-53]. Let us first consider the difficulty (i). The integral over the system on the left-hand side of Eq. (14), i.e., the matrix elements of $\boldsymbol{r}$ with respect to magnetic Bloch functions, can be rewritten as the summation of integrals over the unit cell by using the magnetic Bloch theorem. We have

$$\int_{\Omega} \phi^*_{n\boldsymbol{\kappa}}(\boldsymbol{r}) \boldsymbol{r} \phi_{n'\boldsymbol{\kappa}'}(\boldsymbol{r}) d^3r = \sum_{\boldsymbol{t}_n} \int_{\Omega_{unit}} \phi^*_{n\boldsymbol{\kappa}}(\boldsymbol{r}'+\boldsymbol{t}_n)(\boldsymbol{r}'+\boldsymbol{t}_n) \phi_{n'\boldsymbol{\kappa}'}(\boldsymbol{r}'+\boldsymbol{t}_n) d^3r'$$
$$= N\delta_{\boldsymbol{\kappa},\boldsymbol{\kappa}'} \int_{\Omega_{unit}} \phi^*_{n\boldsymbol{\kappa}}(\boldsymbol{r}') \boldsymbol{r}' \phi_{n'\boldsymbol{\kappa}}(\boldsymbol{r}') d^3r' + \sum_{\boldsymbol{t}_n} \boldsymbol{t}_n e^{i(\boldsymbol{\kappa}'-\boldsymbol{\kappa})\cdot\boldsymbol{t}_n} \int_{\Omega_{unit}} \phi^*_{n\boldsymbol{\kappa}}(\boldsymbol{r}') \phi_{n'\boldsymbol{\kappa}'}(\boldsymbol{r}') d^3r', \tag{B1}$$

where $\boldsymbol{r}' = \boldsymbol{r} - \boldsymbol{t}_n$. The summation in Eq. (B1) is done over the unit cells in the system. The second line of the left-hand side of Eq. (B1) can be obtained by using the magnetic Bloch theorem. It is found that the difficulty (i) does not appear in the second line of the left-hand side. Thus, the difficulty (i) can be avoided by using the magnetic Bloch theorem. As shown in Appendix A, by using the relation $\nabla_{\boldsymbol{\kappa}'} e^{i\boldsymbol{\kappa}'\cdot\boldsymbol{r}} = i\boldsymbol{r} e^{i\boldsymbol{\kappa}'\cdot\boldsymbol{r}}$ in addition to the magnetic Bloch theorem, the matrix elements of $\boldsymbol{r}$ calculated as Eq. (14);

$$\int_{\Omega} \phi^*_{n\boldsymbol{\kappa}}(\boldsymbol{r}) \boldsymbol{r} \phi_{n'\boldsymbol{\kappa}'}(\boldsymbol{r}) d^3r = \delta_{n,n'} \frac{1}{i} \int_{\Omega} \phi^*_{n\boldsymbol{\kappa}}(\boldsymbol{r}) \nabla_{\boldsymbol{\kappa}'} \phi_{n\boldsymbol{\kappa}'}(\boldsymbol{r}) d^3r + \boldsymbol{X}_{nn'}(\boldsymbol{\kappa}) \delta_{\boldsymbol{\kappa},\boldsymbol{\kappa}'}. \tag{14}$$

Similar expressions of the matrix elements of $\boldsymbol{r}$ have been derived by means of the Bloch theorem or magnetic Bloch theorem in previous works [2-4, 7, 12, 49-53]. Such expressions have been widely used in discussions of the anomalous velocity [2-4, 7, 12, 49-53].

Next, we consider the difficulty (ii). Let us consider the case where the origin is shifted by $\boldsymbol{a}$. We denote the position vector after shifting the origin as $\bar{\boldsymbol{r}}$ $(=\boldsymbol{r}-\boldsymbol{a})$. Also, suppose that the periodic potential of the crystal and vector potential after shifting the origin are $\bar{V}(\bar{\boldsymbol{r}})$ and $\bar{\boldsymbol{A}}(\bar{\boldsymbol{r}})$, respectively. We have the following relations:

$$\bar{V}(\boldsymbol{r}-\boldsymbol{a})=V(\boldsymbol{r}), \tag{B2}$$

$$\bar{\boldsymbol{A}}(\boldsymbol{r}-\boldsymbol{a})=\boldsymbol{A}(\boldsymbol{r})+\nabla\chi_{\boldsymbol{a}}(\boldsymbol{r}), \tag{B3}$$

where $\chi_{\boldsymbol{a}}(\boldsymbol{r})$ denotes the gauge transformation function. Note that this gauge transformation comes from the fact that $\boldsymbol{B}=\bar{\nabla}\times\bar{\boldsymbol{A}}(\bar{\boldsymbol{r}})=\nabla\times\boldsymbol{A}(\boldsymbol{r})$, where $\bar{\nabla}$ denotes the nabla operator with respect to $\bar{\boldsymbol{r}}$. If the magnetic Bloch functions after shifting the origin is denoted by $\bar{\phi}_{n\boldsymbol{\kappa}}(\bar{\boldsymbol{r}})$, then we have

$$\bar{\phi}_{n\boldsymbol{\kappa}}(\boldsymbol{r}-\boldsymbol{a})=\phi_{n\boldsymbol{\kappa}}(\boldsymbol{r})e^{-i\frac{e}{\hbar}\chi_{\boldsymbol{a}}(\boldsymbol{r})}. \tag{B4}$$

By using Eq. (B4), we have

$$\int_{\bar{\Omega}}\bar{\phi}^{*}_{n\boldsymbol{\kappa}}(\bar{\boldsymbol{r}})\bar{\boldsymbol{r}}\bar{\phi}_{n'\boldsymbol{\kappa}'}(\bar{\boldsymbol{r}})d^3\bar{r}=\int_{\Omega}\phi^{*}_{n\boldsymbol{\kappa}}(\boldsymbol{r})\boldsymbol{r}\phi_{n'\boldsymbol{\kappa}'}(\boldsymbol{r})d^3r-\boldsymbol{a}\delta_{n\boldsymbol{\kappa},n'\boldsymbol{\kappa}'}. \tag{B5}$$

Thus, the off-diagonal elements of the position operator do not change, while the diagonal elements of the position operator shift uniformly corresponding to the shift of the coordinate origin. Therefore, even if the coordinate origin is shifted, the expected values of the position operator before and after shifting the origin indicate a vector representing the same position. This result is reasonable. Using this result, one can say that only the diagonal elements of the Hamiltonian $\hat{H}$ (Eq. (3)) change by $-\boldsymbol{F}\cdot\boldsymbol{a}$ if the coordinate origin is shifted by $\boldsymbol{a}$. Therefore, the resultant wavefunctions before and after shifting the coordinate origin are related to the gauge transformation such as Eq. (B4). In other words, shifting the coordinate origin does not alter physical quantities such as velocity or current density. Therefore, we can conclude that the difficulty (ii) is not a problem, and that the use of position operators is not a matter of concern when magnetic Bloch functions are used as basis functions.

**APPENDIX C: Derivations of Eqs. (33) and (34)**

In this Appendix, we shall derive Eqs. (33) and (34). Substituting Eq. (14) into Eq. (19), the first term of Eq. (19) is rewritten by

$$
\begin{aligned}
&\frac{1}{i\hbar}\sum_{\boldsymbol{\kappa}}\sum_{\boldsymbol{\kappa}'}\sum_{\boldsymbol{\kappa}''} a^*(\boldsymbol{\kappa},t)a(\boldsymbol{\kappa}',t)\left\{\boldsymbol{F}\cdot\int_\Omega \phi^*_{n\boldsymbol{\kappa}}(\boldsymbol{r})\boldsymbol{r}\phi_{n\boldsymbol{\kappa}''}(\boldsymbol{r})d^3r\right\}\left\{\int_\Omega \phi^*_{n\boldsymbol{\kappa}''}(\boldsymbol{r})\boldsymbol{r}\phi_{n\boldsymbol{\kappa}'}(\boldsymbol{r})d^3r\right\}\\
&=-\frac{1}{i\hbar}\sum_{\boldsymbol{\kappa}}\sum_{\boldsymbol{\kappa}'}\sum_{\boldsymbol{\kappa}''} a^*(\boldsymbol{\kappa},t)a(\boldsymbol{\kappa}',t)\left[\boldsymbol{F}\cdot\int_\Omega \phi^*_{n\boldsymbol{\kappa}}(\boldsymbol{r})\left\{\nabla_{\boldsymbol{\kappa}''}\phi_{n\boldsymbol{\kappa}''}(\boldsymbol{r})\right\}d^3r\right]\int_\Omega \phi^*_{n\boldsymbol{\kappa}''}(\boldsymbol{r})\left\{\nabla_{\boldsymbol{\kappa}'}\phi_{n\boldsymbol{\kappa}'}(\boldsymbol{r})\right\}d^3r\\
&-\frac{1}{\hbar}\sum_{\boldsymbol{\kappa}}\sum_{\boldsymbol{\kappa}'} a^*(\boldsymbol{\kappa},t)a(\boldsymbol{\kappa}',t)\left[\boldsymbol{F}\cdot\int_\Omega \phi^*_{n\boldsymbol{\kappa}}(\boldsymbol{r})\left\{\nabla_{\boldsymbol{\kappa}'}\phi_{n\boldsymbol{\kappa}'}(\boldsymbol{r})\right\}d^3r\right]\boldsymbol{X}_{nn}(\boldsymbol{\kappa}')\\
&-\frac{1}{\hbar}\sum_{\boldsymbol{\kappa}}\sum_{\boldsymbol{\kappa}'} a^*(\boldsymbol{\kappa},t)a(\boldsymbol{\kappa}',t)\left[\left\{\boldsymbol{F}\cdot\boldsymbol{X}_{nn}(\boldsymbol{\kappa})\right\}\int_\Omega \phi^*_{n\boldsymbol{\kappa}}(\boldsymbol{r})\left\{\nabla_{\boldsymbol{\kappa}'}\phi_{n\boldsymbol{\kappa}'}(\boldsymbol{r})\right\}d^3r\right]\\
&+\frac{1}{i\hbar}\sum_{\boldsymbol{\kappa}} a^*(\boldsymbol{\kappa},t)a(\boldsymbol{\kappa}',t)\left\{\boldsymbol{F}\cdot\boldsymbol{X}_{nn}(\boldsymbol{\kappa})\right\}\boldsymbol{X}_{nn}(\boldsymbol{\kappa}).
\end{aligned}
\tag{C1}
$$

By applying $\vec{F}\cdot\nabla_{\boldsymbol{\kappa}'}$ and $\nabla_{\boldsymbol{\kappa}'}$ to the whole of the second and third terms of Eq. (C1), respectively, and then subtracting extra terms, we can rewrite the second and third terms of Eq. (C1). The resultant expression for the first term of Eq. (19) is given by

$$
\begin{aligned}
&\frac{1}{i\hbar}\sum_{\boldsymbol{\kappa}}\sum_{\boldsymbol{\kappa}'}\sum_{\boldsymbol{\kappa}''} a^*(\boldsymbol{\kappa},t)a(\boldsymbol{\kappa}',t)\left\{\int_\Omega \phi^*_{n\boldsymbol{\kappa}}(\boldsymbol{r})\boldsymbol{r}\phi_{n\boldsymbol{\kappa}''}(\boldsymbol{r})d^3r\right\}\left\{\boldsymbol{F}\cdot\int_\Omega \phi^*_{n\boldsymbol{\kappa}''}(\boldsymbol{r})\boldsymbol{r}\phi_{n\boldsymbol{\kappa}'}(\boldsymbol{r})d^3r\right\}\\
&=-\frac{1}{i\hbar}\sum_{\boldsymbol{\kappa}}\sum_{\boldsymbol{\kappa}'}\sum_{\boldsymbol{\kappa}''} a^*(\boldsymbol{\kappa},t)a(\boldsymbol{\kappa}',t)\left[\boldsymbol{F}\cdot\int_\Omega \phi^*_{n\boldsymbol{\kappa}}(\boldsymbol{r})\left\{\nabla_{\boldsymbol{\kappa}''}\phi_{n\boldsymbol{\kappa}''}(\boldsymbol{r})\right\}d^3r\right]\int_\Omega \phi^*_{n\boldsymbol{\kappa}''}(\boldsymbol{r})\left\{\nabla_{\boldsymbol{\kappa}'}\phi_{n\boldsymbol{\kappa}'}(\boldsymbol{r})\right\}d^3r\\
&-\frac{1}{\hbar}\sum_{\boldsymbol{\kappa}}\sum_{\boldsymbol{\kappa}'} a^*(\boldsymbol{\kappa},t)\int_\Omega \phi^*_{n\boldsymbol{\kappa}}(\boldsymbol{r})(\boldsymbol{F}\cdot\nabla_{\boldsymbol{\kappa}'})\left\{a(\boldsymbol{\kappa}',t)\boldsymbol{X}_{nn}(\boldsymbol{\kappa}')\phi_{n\boldsymbol{\kappa}'}(\boldsymbol{r})\right\}d^3r\\
&+\frac{1}{\hbar}\sum_{\boldsymbol{\kappa}} a^*(\boldsymbol{\kappa},t)\boldsymbol{X}_{nn}(\boldsymbol{\kappa})\left\{\boldsymbol{F}\cdot\nabla_{\boldsymbol{\kappa}}a(\boldsymbol{\kappa},t)\right\}+\frac{1}{\hbar}\sum_{\boldsymbol{\kappa}} a^*(\boldsymbol{\kappa},t)a(\boldsymbol{\kappa},t)\left(\boldsymbol{F}\cdot\nabla_{\boldsymbol{\kappa}}\right)\boldsymbol{X}_{nn}(\boldsymbol{\kappa})\\
&-\frac{1}{\hbar}\sum_{\boldsymbol{\kappa}}\sum_{\boldsymbol{\kappa}'} a^*(\boldsymbol{\kappa},t)\left\{\boldsymbol{F}\cdot\boldsymbol{X}_{nn}(\boldsymbol{\kappa})\right\}\int_\Omega \phi^*_{n\boldsymbol{\kappa}}(\boldsymbol{r})\nabla_{\boldsymbol{\kappa}'}\left\{a(\boldsymbol{\kappa}',t)\phi_{n\boldsymbol{\kappa}'}(\boldsymbol{r})\right\}d^3r\\
&+\frac{1}{\hbar}\sum_{\boldsymbol{\kappa}} a^*(\boldsymbol{\kappa},t)\left\{\boldsymbol{F}\cdot\boldsymbol{X}_{nn}(\boldsymbol{\kappa})\right\}\left\{\nabla_{\boldsymbol{\kappa}}a(\boldsymbol{\kappa},t)\right\}+\frac{1}{i\hbar}\sum_{\boldsymbol{\kappa}} a^*(\boldsymbol{\kappa},t)a(\boldsymbol{\kappa},t)\left\{\boldsymbol{F}\cdot\boldsymbol{X}_{nn}(\boldsymbol{\kappa})\right\}\boldsymbol{X}_{nn}(\boldsymbol{\kappa}).
\end{aligned}
\tag{C2}
$$

In a similar way, the second term of Eq. (19) can be rewritten by

$$
\begin{aligned}
&-\frac{1}{i\hbar}\sum_{\boldsymbol{\kappa}}\sum_{\boldsymbol{\kappa}'}\sum_{\boldsymbol{\kappa}''}a^*(\boldsymbol{\kappa},t)a(\boldsymbol{\kappa}',t)\int_\Omega\phi^*_{n\boldsymbol{\kappa}}(\boldsymbol{r})\boldsymbol{r}\phi_{n\boldsymbol{\kappa}''}(\boldsymbol{r})d^3r\int_\Omega\phi^*_{n\boldsymbol{\kappa}''}(\boldsymbol{r})\left(\boldsymbol{F}\cdot\boldsymbol{r}\right)\phi_{n\boldsymbol{\kappa}'}(\boldsymbol{r})d^3r\\
&=\frac{1}{i\hbar}\sum_{\boldsymbol{\kappa}}\sum_{\boldsymbol{\kappa}'}\sum_{\boldsymbol{\kappa}''}a^*(\boldsymbol{\kappa},t)a(\boldsymbol{\kappa}',t)\int_\Omega\phi^*_{n\boldsymbol{\kappa}}(\boldsymbol{r})\left\{\nabla_{\boldsymbol{\kappa}''}\phi_{n\boldsymbol{\kappa}''}(\boldsymbol{r})\right\}d^3r\left[\boldsymbol{F}\cdot\int_\Omega\phi^*_{n\boldsymbol{\kappa}''}(\boldsymbol{r})\left\{\nabla_{\boldsymbol{\kappa}'}\phi_{n\boldsymbol{\kappa}'}(\boldsymbol{r})\right\}d^3r\right]\\
&+\frac{1}{\hbar}\sum_{\boldsymbol{\kappa}}\sum_{\boldsymbol{\kappa}'}a^*(\boldsymbol{\kappa},t)\int_\Omega\phi^*_{n\boldsymbol{\kappa}}(\boldsymbol{r})\nabla_{\boldsymbol{\kappa}'}\left[a(\boldsymbol{\kappa}',t)\left\{\boldsymbol{F}\cdot\boldsymbol{X}_{nn}(\boldsymbol{\kappa}')\right\}\phi_{n\boldsymbol{\kappa}'}(\boldsymbol{r})\right]d^3r\\
&-\frac{1}{\hbar}\sum_{\boldsymbol{\kappa}}a^*(\boldsymbol{\kappa},t)\left\{\boldsymbol{F}\cdot\boldsymbol{X}_{nn}(\boldsymbol{\kappa})\right\}\nabla_{\boldsymbol{\kappa}}a(\boldsymbol{\kappa},t)-\frac{1}{\hbar}\sum_{\boldsymbol{\kappa}}a^*(\boldsymbol{\kappa},t)a(\boldsymbol{\kappa},t)\nabla_{\boldsymbol{\kappa}}\left\{\boldsymbol{F}\cdot\boldsymbol{X}_{nn}(\boldsymbol{\kappa})\right\}\\
&+\frac{1}{\hbar}\sum_{\boldsymbol{\kappa}}\sum_{\boldsymbol{\kappa}'}a^*(\boldsymbol{\kappa},t)\boldsymbol{X}_{nn}(\boldsymbol{\kappa})\int_\Omega\phi^*_{n\boldsymbol{\kappa}}(\boldsymbol{r})\left(\boldsymbol{F}\cdot\nabla_{\boldsymbol{\kappa}'}\right)\left\{a(\boldsymbol{\kappa}',t)\phi_{n\boldsymbol{\kappa}'}(\boldsymbol{r})\right\}d^3r\\
&-\frac{1}{\hbar}\sum_{\boldsymbol{\kappa}}a^*(\boldsymbol{\kappa},t)\boldsymbol{X}_{nn}(\boldsymbol{\kappa})\left\{\boldsymbol{F}\cdot\nabla_{\boldsymbol{\kappa}}a(\boldsymbol{\kappa},t)\right\}-\frac{1}{i\hbar}\sum_{\boldsymbol{\kappa}}a^*(\boldsymbol{\kappa},t)a(\boldsymbol{\kappa},t)\boldsymbol{X}_{nn}(\boldsymbol{\kappa})\left\{\boldsymbol{F}\cdot\boldsymbol{X}_{nn}(\boldsymbol{\kappa})\right\}.
\end{aligned}
\tag{C3}
$$

Substituting Eqs. (C2) and (C3) into Eq. (19), we have

$$
\begin{aligned}
\boldsymbol{v}^{\mathrm{SB},n}_{\mathrm{ano}}&=-\frac{1}{\hbar}\sum_{\boldsymbol{\kappa}}\left|a(\boldsymbol{\kappa},t)\right|^2\boldsymbol{F}\times\boldsymbol{\Omega}_{nn}(\boldsymbol{\kappa})\\
&-\frac{1}{i\hbar}\sum_{\boldsymbol{\kappa}}\sum_{\boldsymbol{\kappa}'}\sum_{\boldsymbol{\kappa}''}a^*(\boldsymbol{\kappa},t)a(\boldsymbol{\kappa}',t)\left[\int_\Omega\phi^*_{n\boldsymbol{\kappa}}(\boldsymbol{r})\boldsymbol{F}\cdot\left\{\nabla_{\boldsymbol{\kappa}''}\phi_{n\boldsymbol{\kappa}''}(\boldsymbol{r})\right\}d^3r\int_\Omega\phi^*_{n\boldsymbol{\kappa}''}(\boldsymbol{r})\left\{\nabla_{\boldsymbol{\kappa}'}\phi_{n\boldsymbol{\kappa}'}(\boldsymbol{r})\right\}d^3r\right]\\
&+\frac{1}{i\hbar}\sum_{\boldsymbol{\kappa}}\sum_{\boldsymbol{\kappa}'}\sum_{\boldsymbol{\kappa}''}a^*(\boldsymbol{\kappa},t)a(\boldsymbol{\kappa}',t)\left[\int_\Omega\phi^*_{n\boldsymbol{\kappa}}(\boldsymbol{r})\left\{\nabla_{\boldsymbol{\kappa}''}\phi_{n\boldsymbol{\kappa}''}(\boldsymbol{r})\right\}d^3r\int_\Omega\phi^*_{n\boldsymbol{\kappa}''}(\boldsymbol{r})\boldsymbol{F}\cdot\left\{\nabla_{\boldsymbol{\kappa}'}\phi_{n\boldsymbol{\kappa}'}(\boldsymbol{r})\right\}d^3r\right]\\
&-\frac{1}{\hbar}\sum_{\boldsymbol{\kappa}}a^*(\boldsymbol{\kappa},t)\int_\Omega\phi^*_{n\boldsymbol{\kappa}}(\boldsymbol{r})\sum_{\boldsymbol{\kappa}'}(\boldsymbol{F}\cdot\nabla_{\boldsymbol{\kappa}'})\left\{a(\boldsymbol{\kappa}',t)\boldsymbol{X}_{nn}(\boldsymbol{\kappa}')\phi_{n\boldsymbol{\kappa}'}(\boldsymbol{r})\right\}d^3r\\
&-\frac{1}{\hbar}\sum_{\boldsymbol{\kappa}}\left\{\boldsymbol{F}\cdot\boldsymbol{X}_{nn}(\boldsymbol{\kappa})\right\}a^*(\boldsymbol{\kappa},t)\int_\Omega\phi^*_{n\boldsymbol{\kappa}}(\boldsymbol{r})\sum_{\boldsymbol{\kappa}'}\nabla_{\boldsymbol{\kappa}'}\left\{a(\boldsymbol{\kappa}',t)\phi_{n\boldsymbol{\kappa}'}(\boldsymbol{r})\right\}d^3r\\
&+\frac{1}{\hbar}\sum_{\boldsymbol{\kappa}}a^*(\boldsymbol{\kappa},t)\int_\Omega\phi^*_{n\boldsymbol{\kappa}}(\boldsymbol{r})\sum_{\boldsymbol{\kappa}'}\nabla_{\boldsymbol{\kappa}'}\cdot\left[\phi_{n\boldsymbol{\kappa}'}(\boldsymbol{r})\left\{\boldsymbol{F}\cdot\boldsymbol{X}_{nn}(\boldsymbol{\kappa}')\right\}a(\boldsymbol{\kappa}',t)\right]d^3r\\
&+\frac{1}{\hbar}\sum_{\boldsymbol{\kappa}}a^*(\boldsymbol{\kappa},t)\boldsymbol{X}_{nn}(\boldsymbol{\kappa})\int_\Omega\phi^*_{n\boldsymbol{\kappa}}(\boldsymbol{r})\sum_{\boldsymbol{\kappa}'}\nabla_{\boldsymbol{\kappa}'}\left\{\boldsymbol{F}\phi_{n\boldsymbol{\kappa}'}(\boldsymbol{r})a(\boldsymbol{\kappa}',t)\right\}d^3r.
\end{aligned}
\tag{C4}
$$

Next, let us consider the second and third terms of Eq. (C4). The second term of Eq. (C4) can be rewritten as follows:

$$
\begin{aligned}
&-\frac{1}{i\hbar}\sum_{\boldsymbol{\kappa}}\sum_{\boldsymbol{\kappa}'}\sum_{\boldsymbol{\kappa}''} a^*(\boldsymbol{\kappa},t)a(\boldsymbol{\kappa}',t)\left[\int_\Omega \phi^*_{n\boldsymbol{\kappa}}(\boldsymbol{r})\boldsymbol{F}\cdot\left\{\nabla_{\boldsymbol{\kappa}''}\phi_{n\boldsymbol{\kappa}''}(\boldsymbol{r})\right\}d^3r\right]\int_\Omega \phi^*_{n\boldsymbol{\kappa}''}(\boldsymbol{r})\left\{\nabla_{\boldsymbol{\kappa}'}\phi_{n\boldsymbol{\kappa}'}(\boldsymbol{r})\right\}d^3r \\
&=-\frac{1}{i\hbar}\sum_{\boldsymbol{\kappa}}\sum_{\boldsymbol{\kappa}''} a^*(\boldsymbol{\kappa},t)\left[\int_\Omega \phi^*_{n\boldsymbol{\kappa}}(\boldsymbol{r})\boldsymbol{F}\cdot\left\{\nabla_{\boldsymbol{\kappa}''}\phi_{n\boldsymbol{\kappa}''}(\boldsymbol{r})\right\}d^3r\right]\int_\Omega \phi^*_{n\boldsymbol{\kappa}''}(\boldsymbol{r})\sum_{\boldsymbol{\kappa}'} a(\boldsymbol{\kappa}',t)\left\{\nabla_{\boldsymbol{\kappa}'}\phi_{n\boldsymbol{\kappa}'}(\boldsymbol{r})\right\}d^3r \\
&=-\frac{1}{i\hbar}\sum_{\boldsymbol{\kappa}}\sum_{\boldsymbol{\kappa}''} a^*(\boldsymbol{\kappa},t)\left[\int_\Omega \phi^*_{n\boldsymbol{\kappa}}(\boldsymbol{r})\boldsymbol{F}\cdot\left\{\nabla_{\boldsymbol{\kappa}''}\phi_{n\boldsymbol{\kappa}''}(\boldsymbol{r})\right\}d^3r\right]\int_\Omega \phi^*_{n\boldsymbol{\kappa}''}(\boldsymbol{r})\sum_{\boldsymbol{\kappa}'} \nabla_{\boldsymbol{\kappa}'}\left\{a(\boldsymbol{\kappa}',t)\phi_{n\boldsymbol{\kappa}'}(\boldsymbol{r})\right\}d^3r \\
&+\frac{1}{i\hbar}\sum_{\boldsymbol{\kappa}} a^*(\boldsymbol{\kappa},t)\int_\Omega \phi^*_{n\boldsymbol{\kappa}}(\boldsymbol{r})\sum_{\boldsymbol{\kappa}'}\left[\left(\boldsymbol{F}\cdot\nabla_{\boldsymbol{\kappa}'}\right)\phi_{n\boldsymbol{\kappa}'}(\boldsymbol{r})\right]\nabla_{\boldsymbol{\kappa}'}a(\boldsymbol{\kappa}',t)d^3r \qquad \text{(C5)} \\
&=-\frac{1}{i\hbar}\sum_{\boldsymbol{\kappa}}\sum_{\boldsymbol{\kappa}''} a^*(\boldsymbol{\kappa},t)\left[\int_\Omega \phi^*_{n\boldsymbol{\kappa}}(\boldsymbol{r})\boldsymbol{F}\cdot\left\{\nabla_{\boldsymbol{\kappa}''}\phi_{n\boldsymbol{\kappa}''}(\boldsymbol{r})\right\}d^3r\right]\int_\Omega \phi^*_{n\boldsymbol{\kappa}''}(\boldsymbol{r})\sum_{\boldsymbol{\kappa}'} \nabla_{\boldsymbol{\kappa}'}\left\{a(\boldsymbol{\kappa}',t)\phi_{n\boldsymbol{\kappa}'}(\boldsymbol{r})\right\}d^3r \\
&+\frac{1}{i\hbar}\sum_{\boldsymbol{\kappa}} a^*(\boldsymbol{\kappa},t)\int_\Omega \phi^*_{n\boldsymbol{\kappa}}(\boldsymbol{r})\sum_{\boldsymbol{\kappa}'}\left(\boldsymbol{F}\cdot\nabla_{\boldsymbol{\kappa}'}\right)\left\{\phi_{n\boldsymbol{\kappa}'}(\boldsymbol{r})\nabla_{\boldsymbol{\kappa}'}a(\boldsymbol{\kappa}',t)\right\}d^3r -\frac{1}{i\hbar}\sum_{\boldsymbol{\kappa}} a^*(\boldsymbol{\kappa},t)\left(\boldsymbol{F}\cdot\nabla_{\boldsymbol{\kappa}}\right)\left\{\nabla_{\boldsymbol{\kappa}}a(\boldsymbol{\kappa},t)\right\}.
\end{aligned}
$$

In a similar way, the third term of Eq. (C4) can be rewritten as

$$
\begin{aligned}
&\frac{1}{i\hbar}\sum_{\boldsymbol{\kappa}}\sum_{\boldsymbol{\kappa}'}\sum_{\boldsymbol{\kappa}''} a^*(\boldsymbol{\kappa},t)a(\boldsymbol{\kappa}',t)\left[\int_\Omega \phi^*_{n\boldsymbol{\kappa}}(\boldsymbol{r})\left\{\nabla_{\boldsymbol{\kappa}''}\phi_{n\boldsymbol{\kappa}''}(\boldsymbol{r})\right\}d^3r\int_\Omega \phi^*_{n\boldsymbol{\kappa}''}(\boldsymbol{r})\boldsymbol{F}\cdot\left\{\nabla_{\boldsymbol{\kappa}'}\phi_{n\boldsymbol{\kappa}'}(\boldsymbol{r})\right\}d^3r\right] \\
&=\frac{1}{i\hbar}\sum_{\boldsymbol{\kappa}}\sum_{\boldsymbol{\kappa}''} a^*(\boldsymbol{\kappa},t)\left[\int_\Omega \phi^*_{\alpha\boldsymbol{\kappa}''}(\boldsymbol{r})\sum_{\boldsymbol{\kappa}'}\boldsymbol{F}\cdot\nabla_{\boldsymbol{\kappa}'}\left\{a(\boldsymbol{\kappa}',t)\phi_{n\boldsymbol{\kappa}'}(\boldsymbol{r})\right\}d^3r\right]\int_\Omega \phi^*_{n\boldsymbol{\kappa}}(\boldsymbol{r})\left\{\nabla_{\boldsymbol{\kappa}''}\phi_{n\boldsymbol{\kappa}''}(\boldsymbol{r})\right\}d^3r \qquad \text{(C6)} \\
&-\frac{1}{i\hbar}\sum_{\boldsymbol{\kappa}} a^*(\boldsymbol{\kappa},t)\int_\Omega \phi^*_{n\boldsymbol{\kappa}}(\boldsymbol{r})\sum_{\boldsymbol{\kappa}'}\nabla_{\boldsymbol{\kappa}'}\left\{\phi_{n\boldsymbol{\kappa}'}(\boldsymbol{r})\left(\boldsymbol{F}\cdot\nabla_{\boldsymbol{\kappa}'}\right)a(\boldsymbol{\kappa}',t)\right\}d^3r +\frac{1}{i\hbar}\sum_{\boldsymbol{\kappa}} a^*(\boldsymbol{\kappa},t)\left(\boldsymbol{F}\cdot\nabla_{\boldsymbol{\kappa}}\right)\left\{\nabla_{\boldsymbol{\kappa}}a(\boldsymbol{\kappa},t)\right\}.
\end{aligned}
$$

Substituting (C5) and (C6) into (C4), we finally get

$$
\boldsymbol{v}^{\mathrm{SB},n}_{\mathrm{ano}} = -\frac{1}{\hbar}\sum_{\boldsymbol{\kappa}}\left|a(\boldsymbol{\kappa},t)\right|^2 \boldsymbol{F}\times\boldsymbol{\Omega}_{nn}(\boldsymbol{\kappa}) + \sum_{\boldsymbol{\kappa}}\boldsymbol{v}^{\mathrm{SB},n}_{\mathrm{another}}(\boldsymbol{\kappa}) \qquad (34)
$$

with

$$
\begin{aligned}
\boldsymbol{v}_{\text{another}}^{\text{SB},n}(\boldsymbol{\kappa}) = & \frac{1}{i\hbar} a^*(\boldsymbol{\kappa},t) \sum_{\boldsymbol{\kappa}''} \int_\Omega \phi_{n\boldsymbol{\kappa}}^*(\boldsymbol{r}) \left\{ \nabla_{\boldsymbol{\kappa}''} \phi_{n\boldsymbol{\kappa}''}(\boldsymbol{r}) \right\} d^3r \int_\Omega \phi_{n\boldsymbol{\kappa}''}^*(\boldsymbol{r}) \boldsymbol{F} \cdot \sum_{\boldsymbol{\kappa}'} \nabla_{\boldsymbol{\kappa}'} \left\{ a(\boldsymbol{\kappa}',t) \phi_{n\boldsymbol{\kappa}'}(\boldsymbol{r}) \right\} d^3r \\
& - \frac{1}{i\hbar} a^*(\boldsymbol{\kappa},t) \sum_{\boldsymbol{\kappa}''} \int_\Omega \phi_{n\boldsymbol{\kappa}}^*(\boldsymbol{r}) \boldsymbol{F} \cdot \left\{ \nabla_{\boldsymbol{\kappa}''} \phi_{n\boldsymbol{\kappa}''}(\boldsymbol{r}) \right\} d^3r \int_\Omega \phi_{n\boldsymbol{\kappa}''}^*(\boldsymbol{r}) \sum_{\boldsymbol{\kappa}'} \nabla_{\boldsymbol{\kappa}'} \left\{ a(\boldsymbol{\kappa}',t) \phi_{n\boldsymbol{\kappa}'}(\boldsymbol{r}) \right\} d^3r \\
& + \frac{1}{i\hbar} a^*(\boldsymbol{\kappa},t) \int_\Omega \phi_{n\boldsymbol{\kappa}}^*(\boldsymbol{r}) \sum_{\boldsymbol{\kappa}'} (\boldsymbol{F} \cdot \nabla_{\boldsymbol{\kappa}'}) \left[ \phi_{n\boldsymbol{\kappa}'}(\boldsymbol{r}) \left\{ \nabla_{\boldsymbol{\kappa}'} a(\boldsymbol{\kappa}',t) \right\} \right] d^3r \\
& - \frac{1}{i\hbar} a^*(\boldsymbol{\kappa},t) \int_\Omega \phi_{n\boldsymbol{\kappa}}^*(\boldsymbol{r}) \sum_{\boldsymbol{\kappa}'} \nabla_{\boldsymbol{\kappa}'} \left\{ \phi_{n\boldsymbol{\kappa}'}(\boldsymbol{r}) (\boldsymbol{F} \cdot \nabla_{\boldsymbol{\kappa}'}) a(\boldsymbol{\kappa}',t) \right\} d^3r \\
& - \frac{1}{\hbar} a^*(\boldsymbol{\kappa},t) \int_\Omega \phi_{n\boldsymbol{\kappa}}^*(\boldsymbol{r}) \sum_{\boldsymbol{\kappa}'} (\boldsymbol{F} \cdot \nabla_{\boldsymbol{\kappa}'}) \left\{ a(\boldsymbol{\kappa}',t) \boldsymbol{X}_{nn}(\boldsymbol{\kappa}') \phi_{n\boldsymbol{\kappa}'}(\boldsymbol{r}) \right\} d^3r \\
& - \frac{1}{\hbar} a^*(\boldsymbol{\kappa},t) \left\{ \boldsymbol{F} \cdot \boldsymbol{X}_{nn}(\boldsymbol{\kappa}) \right\} \int_\Omega \phi_{n\boldsymbol{\kappa}}^*(\boldsymbol{r}) \sum_{\boldsymbol{\kappa}'} \nabla_{\boldsymbol{\kappa}'} \left\{ a(\boldsymbol{\kappa}',t) \phi_{n\boldsymbol{\kappa}'}(\boldsymbol{r}) \right\} d^3r \\
& + \frac{1}{\hbar} a^*(\boldsymbol{\kappa},t) \int_\Omega \phi_{n\boldsymbol{\kappa}}^*(\boldsymbol{r}) \sum_{\boldsymbol{\kappa}'} \nabla_{\boldsymbol{\kappa}'} \left[ \phi_{n\boldsymbol{\kappa}'}(\boldsymbol{r}) \left\{ \boldsymbol{F} \cdot \boldsymbol{X}_n(\boldsymbol{\kappa}') \right\} a(\boldsymbol{\kappa}',t) \right] d^3r \\
& + \frac{1}{\hbar} a^*(\boldsymbol{\kappa},t) \boldsymbol{X}_n(\boldsymbol{\kappa}) \int_\Omega \phi_{n\boldsymbol{\kappa}}^*(\boldsymbol{r}) \sum_{\boldsymbol{\kappa}'} \boldsymbol{F} \cdot \nabla_{\boldsymbol{\kappa}'} \left\{ \phi_{n\boldsymbol{\kappa}'}(\boldsymbol{r}) a(\boldsymbol{\kappa}',t) \right\} d^3r.
\end{aligned} \tag{C7}
$$

Each term of the right-hand side of Eq. (C7) contains the summation over $\boldsymbol{\kappa}'$. If we replace the summation with respect to $\boldsymbol{\kappa}'$ by the volume integral with respect to $\boldsymbol{\kappa}'$, then we have

$$
\begin{aligned}
\boldsymbol{v}_{\text{another}}^{\text{SB},n}(\boldsymbol{\kappa}) = & \frac{1}{i\hbar} a^*(\boldsymbol{\kappa},t) \sum_{\boldsymbol{\kappa}''} \int_\Omega \phi_{n\boldsymbol{\kappa}}^*(\boldsymbol{r}) \left\{ \nabla_{\boldsymbol{\kappa}''} \phi_{n\boldsymbol{\kappa}''}(\boldsymbol{r}) \right\} d^3r \int_\Omega \phi_{n\boldsymbol{\kappa}''}^*(\boldsymbol{r}) \boldsymbol{F} \cdot \left[ \frac{V}{(2\pi)^3} \int_{V_{MBZ}} \nabla_{\boldsymbol{\kappa}'} \left\{ a(\boldsymbol{\kappa}',t) \phi_{n\boldsymbol{\kappa}'}(\boldsymbol{r}) \right\} d^3\kappa' \right] d^3r \\
& - \frac{1}{i\hbar} a^*(\boldsymbol{\kappa},t) \sum_{\boldsymbol{\kappa}''} \int_\Omega \phi_{n\boldsymbol{\kappa}}^*(\boldsymbol{r}) \boldsymbol{F} \cdot \left\{ \nabla_{\boldsymbol{\kappa}''} \phi_{n\boldsymbol{\kappa}''}(\boldsymbol{r}) \right\} d^3r \int_\Omega \phi_{n\boldsymbol{\kappa}''}^*(\boldsymbol{r}) \left[ \frac{V}{(2\pi)^3} \int_{V_{MBZ}} \nabla_{\boldsymbol{\kappa}'} \left\{ a(\boldsymbol{\kappa}',t) \phi_{n\boldsymbol{\kappa}'}(\boldsymbol{r}) \right\} d^3\kappa' \right] d^3r \\
& + \frac{1}{i\hbar} a^*(\boldsymbol{\kappa},t) \int_\Omega \phi_{n\boldsymbol{\kappa}}^*(\boldsymbol{r}) \left[ \frac{V}{(2\pi)^3} \int_{V_{MBZ}} (\boldsymbol{F} \cdot \nabla_{\boldsymbol{\kappa}'}) \left[ \phi_{n\boldsymbol{\kappa}'}(\boldsymbol{r}) \left\{ \nabla_{\boldsymbol{\kappa}'} a(\boldsymbol{\kappa}',t) \right\} \right] d^3\kappa' \right] d^3r \\
& - \frac{1}{i\hbar} a^*(\boldsymbol{\kappa},t) \int_\Omega \phi_{n\boldsymbol{\kappa}}^*(\boldsymbol{r}) \left[ \frac{V}{(2\pi)^3} \int_{V_{MBZ}} \nabla_{\boldsymbol{\kappa}'} \left\{ \phi_{n\boldsymbol{\kappa}'}(\boldsymbol{r}) (\boldsymbol{F} \cdot \nabla_{\boldsymbol{\kappa}'}) a(\boldsymbol{\kappa}',t) \right\} d^3\kappa' \right] d^3r \\
& - \frac{1}{\hbar} a^*(\boldsymbol{\kappa},t) \int_\Omega \phi_{n\boldsymbol{\kappa}}^*(\boldsymbol{r}) \left[ \frac{V}{(2\pi)^3} \int_{V_{MBZ}} (\boldsymbol{F} \cdot \nabla_{\boldsymbol{\kappa}'}) \left\{ a(\boldsymbol{\kappa}',t) \boldsymbol{X}_{nn}(\boldsymbol{\kappa}') \phi_{n\boldsymbol{\kappa}'}(\boldsymbol{r}) \right\} d^3\kappa' \right] d^3r \\
& - \frac{1}{\hbar} a^*(\boldsymbol{\kappa},t) \left\{ \boldsymbol{F} \cdot \boldsymbol{X}_{nn}(\boldsymbol{\kappa}) \right\} \int_\Omega \phi_{n\boldsymbol{\kappa}}^*(\boldsymbol{r}) \left[ \frac{V}{(2\pi)^3} \int_{V_{MBZ}} \nabla_{\boldsymbol{\kappa}'} \left\{ a(\boldsymbol{\kappa}',t) \phi_{n\boldsymbol{\kappa}'}(\boldsymbol{r}) \right\} d^3\kappa' \right] d^3r \\
& + \frac{1}{\hbar} a^*(\boldsymbol{\kappa},t) \int_\Omega \phi_{n\boldsymbol{\kappa}}^*(\boldsymbol{r}) \left[ \frac{V}{(2\pi)^3} \int_{V_{MBZ}} \nabla_{\boldsymbol{\kappa}'} \left[ \phi_{n\boldsymbol{\kappa}'}(\boldsymbol{r}) \left\{ \boldsymbol{F} \cdot \boldsymbol{X}_n(\boldsymbol{\kappa}') \right\} a(\boldsymbol{\kappa}',t) \right] d^3\kappa' \right] d^3r \\
& + \frac{1}{\hbar} a^*(\boldsymbol{\kappa},t) \boldsymbol{X}_{nn}(\boldsymbol{\kappa}) \int_\Omega \phi_{n\boldsymbol{\kappa}}^*(\boldsymbol{r}) \boldsymbol{F} \cdot \left[ \frac{V}{(2\pi)^3} \int_{V_{MBZ}} \nabla_{\boldsymbol{\kappa}'} \left\{ a(\boldsymbol{\kappa}',t) \phi_{n\boldsymbol{\kappa}'}(\boldsymbol{r}) \right\} d^3\kappa' \right] d^3r,
\end{aligned} \tag{C8}
$$

where $V_{MBZ}$ denotes the region of the MBZ. As mentioned in Sec. III, the phase of the magnetic

Bloch function is shown to be discontinuous at certain boundaries (see, Fig. 1). Volume integrals with respect to $\boldsymbol{\kappa}'$ in Eq. (C8) are divided into two integrals within $V_I$ and $V_{II}$. If the divergence theorem is used, then volume integrals within $V_I$ and $V_{II}$ are changed to the surface integrals over the surface of $V_I$ and $V_{II}$, respectively. The surface integral over the surface of $V_{II}$ is given by the surface integral over the surface of $V_{MBZ}$ subtracted by the surface integral over $S_b$. Supposing that the surface integral over the surface of $V_{MBZ}$ vanishes due to the periodicity of the MBZ, and using the integral formula that is given in Appendix D, we finally get Eq. (34).

In order to understand another type of anomalous velocity $\boldsymbol{v}_{\text{another}}^{\text{SB},n}(\boldsymbol{\kappa})$, we rewrite Eq. (19) by using the vector triple product expansion $\boldsymbol{A}\times(\boldsymbol{B}\times\boldsymbol{C})=\boldsymbol{B}(\boldsymbol{A}\cdot\boldsymbol{C})-\boldsymbol{C}(\boldsymbol{A}\cdot\boldsymbol{B})$ and using Eq. (14). The resultant expression for $\boldsymbol{v}_{\text{another}}^{\text{SB},n}(\boldsymbol{\kappa})$ is given by

$$\begin{aligned}\boldsymbol{v}_{\text{another}}^{\text{SB},n}(\boldsymbol{\kappa}) = \frac{1}{\hbar}\boldsymbol{F}\times\Big[&\left|a(\boldsymbol{\kappa},t)\right|^2\boldsymbol{\Omega}_{nn}(\boldsymbol{\kappa})\\ &+\sum_{\boldsymbol{\kappa}'}a^*(\boldsymbol{\kappa},t)a(\boldsymbol{\kappa}',t)\left\{\boldsymbol{X}_{nn}(\boldsymbol{\kappa})-\boldsymbol{X}_{nn}(\boldsymbol{\kappa}')\right\}\times\int_\Omega\phi_{n\boldsymbol{\kappa}}^*(\boldsymbol{r})\left\{\nabla_{\boldsymbol{\kappa}'}\phi_{n\boldsymbol{\kappa}'}(\boldsymbol{r})\right\}d^3r \\ &-\frac{1}{i}\sum_{\boldsymbol{\kappa}'}\sum_{\boldsymbol{\kappa}''}a^*(\boldsymbol{\kappa},t)a(\boldsymbol{\kappa}',t)\int_\Omega\phi_{n\boldsymbol{\kappa}''}^*(\boldsymbol{r})\left\{\nabla_{\boldsymbol{\kappa}'}\phi_{n\boldsymbol{\kappa}'}(\boldsymbol{r})\right\}d^3r\times\int_\Omega\phi_{n\boldsymbol{\kappa}}^*(\boldsymbol{r}')\left\{\nabla_{\boldsymbol{\kappa}''}\phi_{n\boldsymbol{\kappa}''}(\boldsymbol{r}')\right\}d^3r'\Big].\end{aligned} \tag{C9}$$

Equation (C9) implies that $\boldsymbol{v}_{\text{another}}^{\text{SB},n}(\boldsymbol{\kappa})$ is perpendicular to the electric field, similarly to the anomalous velocity.

## APPENDIX D. Integral Formula used in deriving Eq. (34)

In this appendix, we give a proof for the integral formula used in the derivation of Eq. (34). Suppose that $V$ denotes a volume that has a surface $S$. If $\boldsymbol{F}$ and $\boldsymbol{A}(\boldsymbol{r})$ denote a constant vector and a continuously differentiable vector field, respectively, then we can prove the following integral formula,

$$\begin{aligned}\int_V(\boldsymbol{F}\cdot\nabla)\boldsymbol{A}(\boldsymbol{r})d^3r &= \int_S\boldsymbol{A}(\boldsymbol{r})\left\{\boldsymbol{F}\cdot\boldsymbol{n}(\boldsymbol{r})\right\}dS\\ &=\int_S\boldsymbol{A}(\boldsymbol{r})\left(\boldsymbol{F}\cdot d\boldsymbol{S}\right),\end{aligned} \tag{D1}$$

where $\boldsymbol{n}(\boldsymbol{r})$ is the unit normal vector at each point on the boundary $S$. The proof is as follows. According to the vector formula, we have

$$\left(\boldsymbol{F}\cdot\nabla\right)\boldsymbol{A}(\boldsymbol{r})=\nabla\left\{\boldsymbol{F}\cdot\boldsymbol{A}(\boldsymbol{r})\right\}-\boldsymbol{F}\times\left\{\nabla\times\boldsymbol{A}(\boldsymbol{r})\right\}. \tag{D2}$$

Integrating both sides of Eq. (D2), and using integral theorem related to the divergence theorem, we get

$$\int_V\left(\boldsymbol{F}\cdot\nabla\right)\boldsymbol{A}(\boldsymbol{r})d^3r=\int_S\left\{\boldsymbol{F}\cdot\boldsymbol{A}(\boldsymbol{r})\right\}\boldsymbol{n}(\boldsymbol{r})dS-\int_S\boldsymbol{F}\times\left\{\boldsymbol{n}(\boldsymbol{r})\times\boldsymbol{A}(\boldsymbol{r})\right\}dS\,, \tag{D3}$$

where we used the following integral theorems related to divergence theorem [75]:

$$\begin{aligned}&\int_V\nabla\left\{\boldsymbol{F}\cdot\boldsymbol{A}(\boldsymbol{r})\right\}d^3r=\int_S\left\{\boldsymbol{F}\cdot\boldsymbol{A}(\boldsymbol{r})\right\}\boldsymbol{n}(\boldsymbol{r})dS,\\&\int_V\nabla\times\boldsymbol{A}(\boldsymbol{r})d^3r=\int_S\left\{\boldsymbol{n}(\boldsymbol{r})\times\boldsymbol{A}(\boldsymbol{r})\right\}dS.\end{aligned} \tag{D4}$$

By applying the formula for the vector triple product expansion $\boldsymbol{F}\times\left\{\boldsymbol{n}(\boldsymbol{r})\times\boldsymbol{A}(\boldsymbol{r})\right\}=\left\{\boldsymbol{F}\cdot\boldsymbol{A}(\boldsymbol{r})\right\}\boldsymbol{n}(\boldsymbol{r})-\left\{\boldsymbol{F}\cdot\boldsymbol{n}(\boldsymbol{r})\right\}\boldsymbol{A}(\boldsymbol{r})$ to the second term of Eq. (D3), we finally get Eq. (D1).

Figure caption

Fig. 1: Schematic diagram of the MBZ. The region of the MBZ ($V_{MBZ}$) is supposed to be divided into two regions $V_I$ and $V_{II}$. The boundary between $V_I$ and $V_{II}$ is denoted as $S_b$.